\documentclass[12pt]{article}
\usepackage{hyperref}
\usepackage{mathtools}
\usepackage{bm}
\usepackage{float}
\usepackage{relsize}
\usepackage{array}
\usepackage{fullpage}
\usepackage{graphicx}
\usepackage{psfrag}
\usepackage{epsf}
\usepackage{amsfonts}
\usepackage{amsmath}
\usepackage{pstricks}
\usepackage{color}
\usepackage{setspace}
\usepackage{easybmat}
\usepackage{slashed}
\usepackage{amsmath}
\usepackage{amssymb}
\usepackage{graphicx}
\usepackage{verbatim}
\usepackage{caption}
\usepackage{subcaption}
\usepackage{multirow}
\usepackage{tikz}
\usetikzlibrary{matrix,decorations.pathreplacing,calc}

\allowdisplaybreaks
\def\timesbox{\hbox{$\scriptscriptstyle\times$}}
\def\ant{ {{\lower 1ex  \timesbox} \atop {\raise 1.5ex  \timesbox}}}
\def\f{\frac}
\def\pa{\partial}
\def\non{\nonumber\\}
\def\a{\alpha}
\def\b{\beta}

\def\d{\delta}

\def\l{\lambda}

\def\o{\omega}

\def\s{\sigma}
\def\t{\tau}

\newcommand{\mathsym}[1]{{}}
\newcommand{\unicode}[1]{{}}

\newcommand {\expect}[1]{\left\langle #1 \right\rangle}

\newcommand{\be}{\begin{equation}}
\newcommand{\ee}{\end{equation}}
\newcommand{\beqa}{\begin{eqnarray}}
\newcommand{\eeqa}{\end{eqnarray}}
\newcommand{\bsp}{\begin{split}}
\newcommand{\esp}{\end{split}}
\newcommand{\bgth}{\begin{gather}}
\newcommand{\egth}{\end{gather}}

\newcommand{\tr}{\hbox{tr}}
\newcommand{\hf}{{1\over 2}}

\newcommand{\zt}{\zeta}

\newcommand\scalemath[2]{\scalebox{#1}{\mbox{\ensuremath{\displaystyle #2}}}}

\begin{document}

\title{\textbf{Intersecting $D$-Brane Stacks and Tachyons at Finite Temperature}
\author{
Swarnendu Sarkar\footnote{ssarkar@physics.du.ac.in, swarnen@gmail.com}~ and Varun Sethi\footnote{vsethi@physics.du.ac.in}\\
\small{{\em Department of Physics and Astrophysics,
University of Delhi,}} \\ 
\small{{\em Delhi 110007, India}}\\
\\
}}

\maketitle

\abstract{In \cite{2} and \cite{1} intersecting $D$-branes in flat space were studied at finite temperature in the Yang-Mills approximation. The one-loop correction to the tachyon mass was computed and the critical temperature at which the tachyon becomes massless was obtained numerically. In this paper we extend the computation of one-loop two-point amplitude to the case of intersecting stacks of $D3$-branes in flat space. The motivation for this calculation is to study the strong coupling holographic BCS model proposed in \cite{3} at finite temperature. We show that the analytical results of \cite{2} and \cite{1} can be embedded into this more general setup. The main technicality involved here is keeping track of the extra color factors coming from the unbroken gauge groups. We further discuss the issues involved in the computation of two point amplitude for case of multiple intersecting stacks of branes.}

\newpage

\tableofcontents

\baselineskip=18pt

\section{Introduction}\label{intro}

Holographic studies of phases in strongly coupled systems have been of considerable interest in the recent years. One of the most studied example is that of a system that undergoes transition to a superconducting phase \cite{G}-\cite{Hart}. Most of these studies focus on an effective Landau-Ginsburg type of approach. A microscopic top-down model of a holographic superconductor was proposed in \cite{3}. This model is based on a modification of Witten-Sakai-Sugimoto model of holographic QCD \cite{EW1}, \cite{SS1}. For a partial list of other variants of holographic QCD models and related studies see \cite{KK}-\cite{Dhar}. In \cite{3}, the holographic bulk description is in terms of $N_f$ $D8$ branes in the background generated $N_c$ number of $D4$ branes such that $N_c>>N_f$. Intersecting configuration of $D8$ branes in the bulk is unstable due to the appearance of tachyons in the spectrum. The tachyons show up in the spectrum of open strings with the end points on each brane. This instability is proposed to be the dual of Cooper-pairing instability in strongly coupled BCS superconductors. In \cite{3}, this model was studied at zero temperature when the phase is the superconducting BCS phase. 

Intersecting configurations of branes have been studied by various authors. See for example \cite{BDL}-\cite{Jones}.   
Tachyons are known to appear in the open string spectrum of certain non-BPS configurations of intersecting branes.  It was shown, in \cite{Has1}-\cite{Epp1}, that the unstable system gives way to a smooth brane configuration when the tachyons condense. For other analyses see \cite{Nagaoka:2003zn}-\cite{ohta3}. To study the effect of temperature on the system, a setup of intersecting $D1$ branes was considered in \cite{2}. This is a simplified version of the intersecting configuration of $D8$ branes in the holographic model of \cite{3}. The one-loop tachyon mass was computed and it was further shown that there exists a critical temperature $T_c$ at which the tachyon becomes massless. Thus the brane configuration becomes stable above $T_c$. This is what is expected in a BCS system as well. 

The computation in \cite{2} is done in the Yang-Mills approximation $(\alpha^{'}\rightarrow 0, \theta \rightarrow 0)$, with $\theta/(2\pi\alpha^{'})=q$ fixed. The most important point regarding this calculation is that the theory is ultraviolet finite. One can thus compute the critical temperature in a regularization independent way. In \cite{1}, this computation was extended to the case of intersecting $D2$ and $D3$ branes. In both these works the ultraviolet finiteness of the one-loop amplitude was demonstrated explicitly for completeness and as a check for the correctness of the various combinatoric factors.  

This paper generalizes the computations of \cite{1} to the case of intersecting stacks of $D3$ branes. First, we consider two stacks, each with $M$ 
coincident $D$-branes, intersecting at a non-zero angle, $\theta$. This configuration has $SU(M)\times SU(M)\times U(1)$ symmetry. The number of tachyons in this configuration, which is equal to the number of distinct open strings with their two ends on different $D$-branes, is $2M^2$. These tachyons transform as bi-fundamental representation of $SU(M) \times SU(M)$ or as adjoint of $SU(M)$. The intersecting brane configuration, in Yang-Mills approximation, amounts to a gauge theory with $SU(2M)$ symmetry being broken to $SU(M)\times SU(M) \times U(1) $ by giving an expectation value to a scalar field. We then generalize this setup such that the two stacks have different number, $M_1$ and $M_2$, of $D$-branes but again intersecting at a common angle $\theta$. This time, there are $2M_1M_2$ tachyons in the spectrum. Lastly, we consider an arbitrary number of $D$-brane stacks. In this case, there are tachyons coming from each pair of intersecting stacks of $D$-branes. The holographic model considered in \cite{3} involves only two stacks of intersecting branes. However going beyond this model, the multiple intersecting stacks discussed here is a natural generalization.
The result for the previous cases could have been obtained by directly considering this more general setup of multiple stacks. However we have considered these separately as these cases are technically easier to start with, and further they serve as  alternate ways to check our computation. 

As in the previous works we compute the two-point tachyon amplitude for tachyons appearing in these configurations at finite temperature. For the case of two intersecting stacks, the computations here differ from those in the previous works by the appearance of extra color factors from the unbroken gauge groups. Specifically, for the case of $D3$ branes it is shown that the one-loop amplitude is equal to the amplitude computed in \cite{1} times a color factor. This is demonstrated through the computation of some sample Feynman diagrams. This computation however does not naturally generalize to the case where there are more than two stacks. Some technical details are discussed in the paper.

The tachyon mass as a function of temperature can be calculated from the two point amplitude. The temperature at which this vanishes gives the critical temperature. This has to be done numerically. Unlike in the previous works we do not pursue this numerical computation here. The purpose here is to show how the previous analytical computations generalize to the case of intersecting stacks. The only dimensionful parameter in this theory, apart from the temperature, is $q$. The dependence of $T_c$ on $q$ and the Yang-Mills coupling constant $g$ is discussed towards the end of the paper.

The organization of this paper is as follows. Intersecting configuration with two stacks of $M$ $D$-branes is discussed in section \ref{spectrum}.
In section \ref{gen} we write down the generators of $SU(2M)$ in terms of those of $SU(M)$. The tree-level spectrum for bosons and fermions is analyzed in sections \ref{treenn} and \ref{treennf} respectively. The two point amplitude for tachyon, with some sample computations, is presented in section \ref{amplitude}. Ultraviolet and infrared issues involved in this computation are reviewed and discussed in section \ref{uvir}.
Section \ref{M1M2} is devoted to analyzing the one-loop amplitude for the case of two intersecting stacks containing $M_1$ and $M_2$ $D$-branes. We discuss the generalization to multiple stacks of intersecting branes in section \ref{multiple}. We summarize our results in section \ref{summary}.
In appendix \ref{dimred} we have included the ${\cal N}=4$ SYM action in $4D$. The various eigenfunctions involved in the computation of one-loop amplitude are listed in appendix \ref{eigenfns}.

\section{$SU(2M)\rightarrow SU(M)\times SU(M)\times U(1)$ }\label{spectrum}

In this section we study the intersecting brane configuration consisting of two stacks of $M$ $D3$ branes in flat space. 
We are interested in computing the two point amplitude for the tachyons that arise in the spectrum. We first start by writing down the generators of
$SU(2M)$ in terms of those of $SU(M)$. We further analyze the spectrum for the intersecting configuration following \cite{1} and \cite{2}.
As mentioned in the introduction, we shall do the computations in the Yang-Mills approximation.

\subsection{$SU(2M)$ generators}\label{gen}

With a view to studying the broken symmetry configuration, we first write down the generators $SU(2M)$ in terms of those of $SU(M)$. Let us denote the generators of $SU(M)$ by $\l^{\bm{a}}$ with $\bm{a}=1, \cdots, M^2-1$ and let $\l^0=\f{1}{\sqrt{2M}}\mathbb{I}_{M\times M}$. These matrices satisfy $\tr(\l^0)^2=\tr(\l^{\bm{a}})^2=\f{1}{2}$.

Two intersecting stacks of coincident $M$ number of branes is achieved by turning on expectation value of one of the scalars in (\ref{n4action}) as $qx \f{1}{\sqrt{2}}\left(\l^0\otimes\s^3\right)$. The other generators of $SU(2M)$ which commute with this and hence remain unbroken are
\beqa
\f{1}{\sqrt{2}}\left(\l^{\bm{a}}\otimes\s^0\right)~~~;~~~\f{1}{\sqrt{2}}\left(\l^{\bm{a}}\otimes\s^3\right)
\eeqa

where $\s^0=\mathbb{I}_{2\times2}$ and  $\s^{\bm{i}}$, $\bm{i}=1,2,3$ are Pauli matrices. The unbroken generators of $SU(2M)$ are thus $2M^2-1$ in number.

The broken generators are
 
\beqa
\f{1}{\sqrt{2}}\left(\l^{\bm{a}}\otimes\s^{1,2}\right)~~~;~~~\f{1}{\sqrt{2}}\left(\l^0\otimes\s^{1,2}\right)
\eeqa

which are $2M^2$ in number. So the total number of generators of $SU(2M)$ add up to $4M^2-1$.

\subsection{Tree-level spectrum of Bosons}\label{treenn}

In this section we study here the tree-level spectrum of bosons for the configuration of intersecting stacks of branes in the Yang-Mills approximation. The computations in this section are adapted from that of \cite{1} and differ from the latter by the appearance of extra color 
indices.

The adjoint gauge and scalar fields with the $SU(2M)$ generators defined above are written as 

\beqa
A_{\mu}=A_{\mu}^{ai}\f{1}{\sqrt{2}}\left(\l^a\otimes\s^{{i}}\right)~~~;~~~\Phi_I=\Phi_{I}^{a{i}}\f{1}{\sqrt{2}}\left(\l^a\otimes\s^{{i}}\right)
\eeqa

with $a\equiv(0, \bm{a})$ and ${i}\equiv (0, \bm{i})$.

As mentioned before, the intersecting brane configuration corresponds to setting the background value of one of the scalar fields equal to
$qx$. We choose this scalar field to be $\Phi_1^{03}$. To see the coupling between the fields at the quadratic level let us consider the following
term in the action (\ref{n4action}),

\beqa\label{quadterm}
-2i~ \tr \left(\pa^{\mu}\Phi_1\left[A_{\mu},\Phi_1\right]\right).
\eeqa

Expanding about the background value $\Phi_1=qx \f{1}{\sqrt{2}}\left(\l^0\otimes\s^3\right)$, the resulting terms quadratic in the fluctuations are 

\beqa\label{expand1}
&-&2qi ~\tr\left(x\pa^{\mu}\Phi_1\left[A_{\mu},\f{1}{\sqrt{2}}\left(\l^0\otimes\s^3\right)\right]+\f{1}{\sqrt{2}}\left(\l^0\otimes\s^3\right)\left[A_1,\Phi_1\right]\right)\non
&=& \f{q}{\sqrt{M}}\left(x\pa^{\mu}\Phi_1^{a1}A_{\mu}^{a2}-x\pa^{\mu}\Phi_1^{a2}A_{\mu}^{a1}-\Phi_1^{a1}A_1^{a2}+\Phi_1^{a2}A_1^{a1}\right)
\eeqa

In the following, we shall absorb the factor of $\f{1}{\sqrt{M}}$ appearing in equation (\ref{expand1}) into a re-definition of $q$. This will make it is easier for us to compare the results obtained here to those of \cite{1}. For notational convenience we shall denote the redefined quantity also by $q$.

Following equation (\ref{expand1}), let us define

\beqa\label{multiplet}
\xi^{a}=\left(\begin{array}{c}\Phi^{a1}_1\\A^{a2}_1\\A^{a2}_2\\A^{a2}_3\end{array}\right)~;~\zeta^{a}=\left(\begin{array}{c}\Phi^{a1}_1\\A^{a2}_1\end{array}\right)~~~;~~~\xi^{'a}=\left(\begin{array}{c}\Phi^{a2}_1\\A^{a1}_1\\A^{a1}_2\\A^{a1}_3\end{array}\right)~;~\zeta^{'a}=\left(\begin{array}{c}\Phi^{a2}_1\\A^{a1}_1\end{array}\right)
\eeqa

The full quadratic bosonic part of the action is then,

\beqa\label{actionb}
S_b=\int d^4 z ~\left[\f{1}{2}(\xi^{a})^T{\cal O}_B\xi^{a}+\f{1}{2}(\xi^{'a})^{T}{\cal O}^{'}_B\xi^{'a}+{\cal L}(A^{a0}_{\mu},A^{a3}_{\mu},\Phi_I,\tilde{\Phi}_J)\right].
\eeqa

Identifying  $z^1=x$, the operator ${\cal O}_B$ is

\beqa\label{ob}
{\cal O}_B=\left(\begin{array}{cc}{\cal O}^{11}_B&{\cal O}^{12}_B\\{\cal O}^{21}_B&{\cal O}^{22}_B\end{array}\right),
\eeqa

where

\beqa
{\cal O}^{11}_B=\left(\begin{array}{cc}-\partial^2_0+\partial^2_1+\partial^2_2+\partial^2_3&-2q-qx\partial_1\\-q+qx\partial_1&-\partial_0^2+\partial_2^2+\partial_3^2-q^2x^2\end{array}\right)~~~({\cal O}^{12}_B)^T={\cal O}^{21}_B=\left(\begin{array}{cc}qx\partial_2&-\partial_2\partial_1\\qx\partial_3&-\partial_3\partial_1\end{array}\right)
\eeqa

\beqa
{\cal O}^{22}_B=\left(\begin{array}{cc}-\partial^2_0+\partial^2_1+\partial^2_3-q^2x^2&-\partial_2\partial_3\\-\partial_2\partial_3&-\partial^2_0+\partial^2_1+\partial^2_2-q^2x^2\end{array}\right)
\eeqa
${\cal O}^{'}_B$ can be obtained from ${\cal O}_B$ by replacing $q$ in ${\cal O}_B$ by $-q$. 

The eigenfunctions of ${\cal O}^{11}_B$ have been first worked out in \cite{Has1} and \cite{Epp1}. These were reviewed in \cite{2} where they were further rewritten in terms of Hermite polynomials. The eigenfunctions along with their various properties are listed in the Appendix \ref{eigenfns}.

To study the theory at finite temperature we shall follow the imaginary time formalism (see for example \cite{kapusta}). We thus identify $z^0=-i\tau$ and $(z^2,z^3)\equiv {\bf y}$ where $\tau$ is periodic with period $\beta$ which is the inverse of the temperature $T$. The mode expansions are written as 

\beqa\label{zeta1}
\zeta^{a}(\tau, x, {\bf y})=N^{1/2}\int\f{d^2{\bf k}}{(2\pi\sqrt{q})^2}\sum_{m,n}\left[C^a(m,n,{\bf k})\zeta_n(x)+
\tilde{A}^{a2}_1(m,n,{\bf k})\tilde{\zeta}_n(x)\right]e^{-i(\o_m\tau+{\bf k.y})}
\eeqa

\beqa
\zeta^{a'}(\tau, x, {\bf y})=N^{1/2}\int\f{d^2{\bf k}}{(2\pi\sqrt{q})^2}\sum_{m,n}\left[C^{'a}(m,n,{\bf k})\zeta^{'}_n(x)+
\tilde{A}^{a1}_1(m,n,{\bf k})\tilde{\zeta}^{'}_n(x)\right]e^{-i(\o_m\tau+{\bf k.y})}
\eeqa

where $\zeta_n(x)$, $\tilde{\zeta}_n(x)$, $\zeta^{'}_n(x)$ and $\tilde{\zeta}^{'}_n(x)$ are the eigenfunctions of the operator ${\cal O}^{11}_B$. $N=\sqrt{q}/\beta$, $\omega_m=2\pi m/\beta$ with $m=0,1,2, \cdots$.

Similarly considering the operator ${\cal O}^{22}_B$ we have the following mode expansions

\beqa
A_{2,3}^{a2}=N^{1/2}\int\f{d^2{\bf k}}{(2\pi\sqrt{q})^2}\sum_{m,n}\tilde{A}_{2,3}^{a2}(m,n,{\bf k}){\cal N}^{'}(n)e^{-qx^2/2}H_n(\sqrt{q}x)e^{-i(\o_m\tau+{\bf k.y})}
\eeqa

where $H_n(\sqrt{q}x)$ are Hermite polynomials and the normalization $\mathcal{N}^{'}(n)=1/\sqrt{\sqrt{\pi}2^n n!}$. Further since

\beqa
{\cal O}^{21}_B \zeta_n=0~~~ \mbox{and}~~~ {\cal O}^{21}_B \tilde{\zeta}_n=2(2n-1)\tilde{{\cal N}}(n)\sqrt{q}e^{-qx^2/2}H_{n-1}(\sqrt{q}x)\left(\begin{array}{c}\partial_2\\\partial_3\end{array}\right)
\eeqa

we have

\beqa
&&{\cal O}^{21}_B \zeta^a=N^{1/2}\int\f{d^2{\bf k}}{(2\pi\sqrt{q})^2}\sum_{m,n}\tilde{A}^{a2}_1(m,n,{\bf k})\times \non
&&~~~~~~~~~~~~~~~~~~~~~~~~~~~~~~~\times\left[2(2n-1)\tilde{{\cal N}}(n)\sqrt{q}e^{-qx^2/2}H_{n-1}(\sqrt{q}x)
\left(\begin{array}{c}\partial_2\partial_3\end{array}\right)e^{-i(\o_m\t+{\bf k}.{\bf y})}\right]
\eeqa

Using these, the part of the action involving the modes of the $\xi^a$ multiplet (defined in (\ref{multiplet})) is

\beqa
&&-\f{1}{2qg^2}\int\f{d^2{\bf k}}{(2\pi\sqrt{q})^2}\sum_{m=-\infty,n=0}^{\infty,\infty}\left[\tilde{A}_i^{a2}(m,n,{\bf k})\left(k^2\delta^{ij}-k^ik^j\right)\tilde{A}_j^{a2}(-m,n,-{\bf k})\right.\non
&&~~~~~~~~~~~~~~~~~~~~~~~~~~~~~~~~~~~~~~~~~\left.  +|C^a(m,n,{\bf k})|^2\left(\o_m^2+\lambda_n+{\bf k}^2\right)\right]
\eeqa

where $(i,j=1,2,3)$, $k^2=(\o_m^2+\gamma_n+{\bf k}^2)$, $C^a(-m,n,-{\bf k})=C^{a*}(m,n,{\bf k})$, $k_x=\sqrt{\gamma_n}=\sqrt{(2n+1)q}$ and $\lambda_n=(2n-1)q$.

Similarly for the momentum modes of $\xi^{'a}$, the action is

\beqa
&&-\f{1}{2qg^2}\int\f{d^2{\bf k}}{(2\pi\sqrt{q})^2}\sum_{m=-\infty,n=0}^{\infty,\infty}\left[\tilde{A}_i^{a1}(m,n,{\bf k})\left(k^2\delta^{ij}-k^ik^j\right)\tilde{A}_j^{a1}(-m,n,-{\bf k})\right.\non 
&&~~~~~~~~~~~~~~~~~~~~~~~~~~~~~~~~~~~~~~~~~\left.+|C^{'a}(m,n,{\bf k})|^2\left(\o_m^2+\lambda_n+{\bf k}^2\right)\right]
\eeqa

where $(i,j=1,2,3)$, $k^2=(\o_m^2+\gamma_n+{\bf k}^2)$, $C^{'a}(-m,n,-{\bf k})=C^{'a*}(m,n,{\bf k})$, $k_x=\sqrt{\gamma_n}=\sqrt{(2n+1)q}$ and $\lambda_n=(2n-1)q$.

The fields $C^a(-m,n,-{\bf k})$ and $C^{'a}(-m,n,-{\bf k})$ are tachyonic for $n=0$. They are $2M^2$ in number, which is equal to the number of broken generators of $SU(2M)$. In the following section, we shall analyze the one-loop two point function for these tachyons.

We now write down the mode expansions for the fields contained in ${\cal L}(A^{a0}_{\mu},A^{a3}_{\mu},\Phi_I,\tilde{\Phi}_J)$ of (\ref{actionb}).  The scalar fields with gauge components $(a1,a2)$ can be expanded as

\beqa
\Phi_{I}^{a1,a2}=N^{1/2}\int\f{d^2{\bf k}}{(2\pi\sqrt{q})^2}\sum_{m,n}\Phi_{I}^{a1,a2}(m,n,{\bf k}){\cal N}^{'}(n)e^{-qx^2/2}H_n(\sqrt{q}x)e^{-i(\o_m\tau+{\bf k.y})}
\eeqa

The scalar fields with the gauge component $(a0,a3)$ and the gauge fields $A^{a0}_{i}$ and  $A^{a3}_{i}$ can be expanded using the basis for plane wave as

\beqa
\Phi_{J}^{a0,a3}&=&N^{1/2}\int\f{dk_xd^2{\bf k}}{(2\pi\sqrt{q})^3}\sum_{m}\Phi_{J}^{a0,a3}(m,k)e^{-i(\o_m\tau+k_x x +{\bf k.y})}~~(J=1,2,3)\\
A_{i}^{a0,a3}&=&N^{1/2}\int\f{dk_xd^2{\bf k}}{(2\pi\sqrt{q})^3}\sum_{m}A_{i}^{a0,a3}(m,k)e^{-i(\o_m\tau+k_x x +{\bf k.y})}~~(i=1,2,3)
\eeqa

The corresponding action in terms of these modes is then

\beqa\label{phiAaction}
&&-\f{1}{2qg^2}\int\f{d^2{\bf k}}{(2\pi\sqrt{q})^2}\sum_{\substack{m,n\\a}}\left[|\Phi^{a1}_{I}(m,n,{\bf k})|^2\left(\o_m^2+\gamma_n+{\bf k}^2\right)+|\tilde{\Phi}^{a1}_{J}(m,n,{\bf k})|^2\left(\o_m^2+\gamma_n+{\bf k}^2\right)\right]\non
&&-\f{1}{2qg^2}\int\f{d^2{\bf k}}{(2\pi\sqrt{q})^2}\sum_{\substack{m,n\\a}}\left[|\Phi^{a2}_{I}(m,n,{\bf k})|^2\left(\o_m^2+\gamma_n+{\bf k}^2\right)+|\tilde{\Phi}^{a2}_{J}(m,n,{\bf k})|^2\left(\o_m^2+\gamma_n+{\bf k}^2\right)\right]\non
&&-\f{1}{2qg^2}\int\f{dk_xd^2{\bf k}}{(2\pi\sqrt{q})^3}\sum_{\substack{m\\a}}\left[|\Phi_J^{a0}(m,k)|^2k^2+|\tilde{\Phi}_J^{a0}(m,k)|^2k^2\right.\non
&&~~~~~~~~~~~~~~~~~~~~~~~~~~~~~~~~~~~~~~~~~~\left.+\tilde{A}_i^{a0}(m, k)\left(k^2\delta^{ij}-k^ik^j\right)\tilde{A}_j^{a0}(-m,-k)\right]\non
&&-\f{1}{2qg^2}\int\f{dk_xd^2{\bf k}}{(2\pi\sqrt{q})^3}\sum_{\substack{m\\a}}\left[|\Phi_J^{a3}(m,k)|^2k^2+|\tilde{\Phi}_J^{a3}(m,k)|^2k^2\right.\non
&&~~~~~~~~~~~~~~~~~~~~~~~~~~~~~~~~~~~~~~~~~~\left.+\tilde{A}_i^{a3}(m, k)\left(k^2\delta^{ij}-k^ik^j\right)\tilde{A}_j^{a3}(-m,-k)\right]\non
\eeqa

Here $(I=2,3)$, $(J=1,2,3)$, $(i,j=1,2,3)$ and $k^2=(\o_m^2+k_x^2+{\bf k}^2)$.

\begin{figure}[h]
\begin{center}
\begin{psfrags}
\psfrag{sum}[][]{$M$}
\psfrag{A3}[][]{$A^{(a0,a3)}/\Phi_I^{(a0,a3)}$}
\psfrag{A2}[][]{$A^{(a1,a2)}/\Phi_I^{(a1,a2)}$}
\includegraphics[width= 6cm,angle=0]{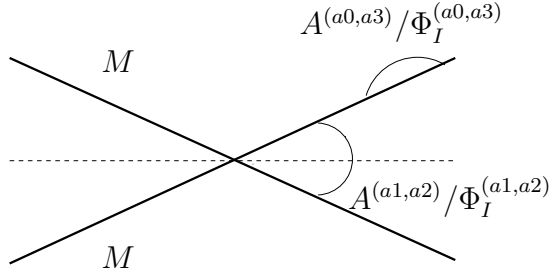}
\caption{Intersecting configuration with two stacks of $M$ $D3$ branes}
\label{intfig1}
\end{psfrags}
\end{center}
\end{figure}

To summarize, the massive modes, including the tachyon, arise from the components to the fields which couple to the background scalar. These are the off-diagonal fields with gauge components $(a1,a2)$ and transform as bi-fundamental under $SU(M)\times SU(M)$ or as adjoint of $SU(M)$. They correspond to open strings stretching from one stack to another (see Figure \ref{intfig1}). The diagonal massless modes 
have end points on the same stack. They correspond to gauge indices $03$ and $(a0,a3)$ transforming under $U(1)$ generated by $\f{1}{\sqrt{2}}\left(\l^0\otimes\s^3\right)$ and as adjoint of $SU(M)$ respectively.

\subsection{Tree-level spectrum : Fermions}\label{treennf}

The quadratic terms involving fermions are obtained from the action (\ref{actionfermion}) by setting the background value of the scalar $\Phi_1=qx \f{1}{\sqrt{2}}\left(\l^0\otimes\s^3\right)$\footnote{The both the $SU(M)$ generators as well fermions in this section have been denoted by $\l$'s. However they are easily distinguished by their index structures.}. The terms after simplification take the following form

\beqa\label{Lf}
{\cal L}^{'}=-\f{i}{2}\left[\bar{\lambda}^{ai}_k\gamma^{\mu}\partial_{\mu}\lambda_k^{ai}+(qx)\left(\bar{\lambda}^{a1}_k\alpha^1_{kl}\lambda^{a2}_l-\bar{\lambda}^{a2}_k\alpha^1_{kl}\lambda^{a1}_l\right)\right]
\eeqa

where $a=0, \cdots, M^2-1$, $i=0,1,2,3$ and $k,l=1,\cdots,4$. We have absorbed a factor of $\f{1}{\sqrt{M}}$ in $q$ in equation (\ref{Lf}). 

Now defining,

\beqa\label{fmultiplets}
\chi_1^{a}=\left(\begin{array}{c}\lambda^{a1}_1\\\lambda^{a2}_4\end{array}\right) ~~~\chi_2^a=\left(\begin{array}{c}\lambda^{a1}_2\\\lambda^{a2}_3\end{array}\right)~~~\chi_3^{a}=\left(\begin{array}{c}\lambda^{a1}_3\\\lambda^{a2}_2\end{array}\right)~~~\chi_4^a=\left(\begin{array}{c}\lambda^{a1}_4\\\lambda^{a2}_1\end{array}\right)
\eeqa

The Lagrangian (\ref{Lf}) can thus be written as 
\beqa
{\cal L}^{'}=-\f{i}{2}\left[\bar{\chi}_1^a {\cal O}_F\chi_1^a+\bar{\chi}_2^a {\cal O}_F\chi_2^a+\bar{\chi}_3^a \tilde{{\cal O}}_F\chi_3^a+\bar{\chi}_4^a \tilde{{\cal O}}_F\chi_4^a+\bar{\lambda}^{a3}_k\gamma^{\mu}\partial_{\mu}\lambda_k^{a3}+\bar{\lambda}^{a0}_k\gamma^{\mu}\partial_{\mu}\lambda_k^{a0}\right].
\eeqa

The operators ${\cal O}_F$ and  $\tilde{{\cal O}}_F$  are given by,

\beqa
{\cal O}_F=\left(\begin{array}{cc}\gamma^{\mu}\partial_{\mu}&qx\\qx&\gamma^{\mu}\partial_{\mu}\end{array}\right) ~~~\tilde{{\cal O}}_F=\left(\begin{array}{cc}\gamma^{\mu}\partial_{\mu}&-qx\\-qx&\gamma^{\mu}\partial_{\mu}\end{array}\right)
\eeqa

We transform $\chi_i \rightarrow U \chi_i$ where the unitary matrix, $U=\left(\begin{array}{cc}1&0\\0&\gamma^1\end{array}\right)$. The operators ${\cal O}_F$ and  $\tilde{{\cal O}}_F$ thus transform into

\beqa
{\cal O}_F\rightarrow\gamma^{i}\partial_{i}\otimes \mathbb{I}_2+\gamma^{1}\otimes{\cal O}_F^x ~~~\mbox{with}~~~{\cal O}_F^x=\left(\begin{array}{cc}\partial_{x}&qx\\-qx&-\partial_{x}\end{array}\right)
\eeqa

\beqa
\tilde{{\cal O}}_F\rightarrow\gamma^{i}\partial_{i}\otimes \mathbb{I}_2+\gamma^{1}\otimes\tilde{{\cal O}}_F^x ~~~\mbox{with}~~~\tilde{{\cal O}}_F^x=\left(\begin{array}{cc}\partial_{x}&-qx\\qx&-\partial_{x}\end{array}\right)
\eeqa

where $i=0,2,3$. The eigenfunctions of the matrix operators have been obtained in \cite{1,2}.These functions are listed in the Appendix \ref{eigenfns}. Using these functions, the mode expansions for the fermions are

\beqa
&&\chi_i^{a}(\tau,x,{\bf y})=N^{3/4}\sum_{n,m}\int\f{d^2{\bf k}}{(2\pi\sqrt{q})^2}\left[\left(\begin{array}{c}\theta_i^a(\o,n,{\bf k})L_n(x)\\\gamma^{1}\theta_i^a(\o,n,{\bf k})R_n(x)\end{array}\right)e^{-i(\o_m\tau+{\bf k.y})}\right.\non
&&~~~~~~~~~~~~~~~~~~~~~~~~~~~~~~~~~~~~~~~~~~~~~~\left.+\left(\begin{array}{c}\theta^{a*}_i(\o,n,{\bf k})L^*_n(x)\\\gamma^{1}\theta^{a*}_i(\o,n,{\bf k})R^*_n(x)\end{array}\right)e^{i(\o_m\tau+{\bf k.y})}\right]
\eeqa

where $\theta^a_i$ are four component fermions and $\o_m=(2m+1)\pi/\beta$ with $m=0,1,2, \cdots$. Further

\beqa
\lambda^{a0,a3}_l(\tau,x,{\bf y})=N^{3/4}\sum_{m}\int\f{dk_xd^2{\bf k}}{(2\pi\sqrt{q})^3}\lambda^{a0,a3}_l(\o,k_x,{\bf k})e^{-i(\o_m\tau+k_x x+{\bf k.y})}
\eeqa

The quadratic action in terms of the momentum modes is then

\beqa
&&S_f=\f{N^{1/2}}{qg^2}\left[\sum_{i,m,n}\int\f{d^2{\bf k}}{(2\pi\sqrt{q})^2}\bar{\theta}_i^{a}(m,n,{\bf k})i\slashed{P}_+\theta_i^{a}(m,n,{\bf k})\right.\non
&&~~~~~~~~~~~~~~\left.+\f{1}{2}\sum_{l,m}\int\f{dk_xd^2{\bf k}}{(2\pi\sqrt{q})^3}\bar{\lambda}^{a0}_l(m,k_x,{\bf k})i\slashed{K}_+\lambda^{a0}_l(m,k_x,{\bf k})\right.\non
&&~~~~~~~~~~~~~~\left.+\f{1}{2}\sum_{l,m}\int\f{dk_xd^2{\bf k}}{(2\pi\sqrt{q})^3}\bar{\lambda}^{a3}_l(m,k_x,{\bf k})i\slashed{K}_+\lambda^{a3}_l(m,k_x,{\bf k})\right]
\eeqa

where $\slashed{P}_+=i\o_m\gamma^0+\sqrt{\lambda^{'}_n}\gamma^1+k_2\gamma^2+k_3\gamma^3$ and $\slashed{K}_+=i\o_m\gamma^0+k_x\gamma^1+k_2\gamma^2+k_3\gamma^3$.

As discussed towards the end of section \ref{treenn} for the case of bosons, the fermions $\theta^a$ correspond to stretched strings from one stack to another. The $\lambda^{a0,a3}$ are the massless ones corresponding to open strings with endpoints on the same stack (see Figure \ref{intfig1}).

\subsection{Sample computations of amplitudes}\label{amplitude}

In this section, we demonstrate the calculations of some of the contributions to the tachyon two-point amplitudes. Our goal is to compare the results with those in \cite{1} where single intersecting branes were considered. It will turn out that the result for the one-loop two-point amplitude here differs from that of \cite{1} by an overall color factor arising from the unbroken gauge group. It will thus be sufficient to compute some of the contributions to the tachyon two-point amplitude and hence deduce that the same color factor arises from all the contributions. The notations here have been kept same as that of \cite{1} with extra color indices wherever involved. This will make it easier for the comparison. 

\begin{figure}[h]
\begin{center}
\begin{psfrags}
\psfrag{c1}[][]{$C^a$}
\psfrag{c2}[][]{$C^c$}
\psfrag{c4}[][]{$(\Phi^{b1}_I,\tilde{\Phi}^{b1}_I)/(\Phi^{b2}_I,\tilde{\Phi}^{b2}_I)$}
\psfrag{c5}[][]{$(\Phi^{b3}_I,\tilde{\Phi}^{b3}_I)/(\Phi^{b0}_I,\tilde{\Phi}^{b0}_I)$}
\psfrag{v2}[][]{$V^{2}_2$}
\psfrag{v3}[][]{$V^{3}_2$}
\psfrag{(I)}[][]{(A)}
\psfrag{(II)}[][]{(B)}
\includegraphics[width= 10cm,angle=0]{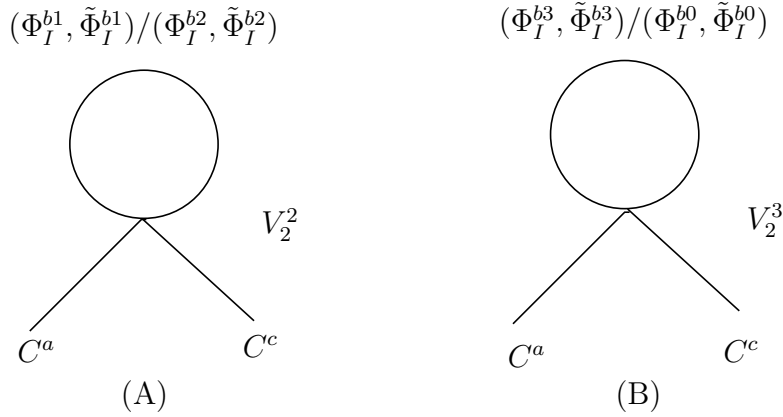}
\caption{Tachyon amplitudes with bosonic 4-point vertices}
\label{tachampboson4pt}
\end{psfrags}
%\label{bosfourfeyn}
\end{center}
\end{figure}

First, we consider tachyon two-point amplitudes with bosonic four-point vertices. Consider the terms $ \hf \tr[\Phi_I,\Phi_J]^2 $, $ \hf \tr[\Phi_I,\tilde{\Phi}_J]^2 $ ,$\tr[A_{\mu},\Phi_I]^2$ and $\tr[A_{\mu},\tilde{\Phi}_I]^2$ in the action, equation (\ref{actionboson}). These contribute to the amplitudes shown in Figure \ref{tachampboson4pt} with tachyons, $C^a,C^c$ being the fields appearing as the coefficients of $\zt_n(x)$ in the mode expansion of $\Phi^{a1}_1$ fields (\ref{zeta1}) for $n=0$, in the multiplet $\zt^a(x)$.

We first illustrate the computation of the vertex resulting from the $ \hf \tr[\Phi_I,\Phi_J]^2 $ term.
With the fields written using $SU(2M)$ generators, this term can be expanded as
\beqa
\hf \tr\left[\left\{\Phi^{ai}_I \f{1}{\sqrt{2}}(\lambda^a \otimes \sigma^i),\Phi^{bj}_J \f{1}{\sqrt{2}}(\lambda^b \otimes \sigma^j)\right\}\left\{\Phi^{ck}_I \f{1}{\sqrt{2}}(\lambda^c \otimes \sigma^k),\Phi^{dl}_I \f{1}{\sqrt{2}}(\lambda^d \otimes \sigma^l)\right\}\right]. \nonumber
\eeqa
We make the choice $ i=k=1$, $I = 1$ for two tachyons on external legs. Simplifying the commutator gives
\beqa
\Phi^{a1}_1
\Phi^{bj}_J
\Phi^{c1}_1
\Phi^{dl}_J
&\tr& \left\{\hf\left([\lambda^a,\lambda^b] \otimes [\sigma^1,\sigma^j] + \lambda^b\lambda^a \otimes [\sigma^1,\sigma^j] + [\lambda^a,\lambda^b] \otimes \sigma^j\sigma^1\right)\right\}\nonumber\\
&\times& \left\{\hf\left([\lambda^c,\lambda^d] \otimes [\sigma^1,\sigma^l] + \lambda^d\lambda^c \otimes [\sigma^1,\sigma^l] + [\lambda^c,\lambda^d] \otimes \sigma^l\sigma^1\right)\right\}
\eeqa

Writing the fields in Fourier modes and further simplifying the commutator and product of Pauli matrices gives

\beqa\label{Sbterm1}
&-&\hf(\epsilon^{1ji'}\epsilon^{1li'})
N^{1/2}\int\f{d^2{\bf k}}{(2\pi\sqrt{q})^2}\sum_{m,n}\left[C^a(m,n,{\bf k})\phi_n(x)+
\tilde{A}^{2a}_1(m,n,{\bf k})\tilde{\phi}_n(x)\right]e^{-i(\o_m\tau+{\bf k.y})}\nonumber\\
&\times& N^{1/2}\int\f{d^2{\bf k}}{(2\pi\sqrt{q})^2}\sum_{m,n}\Phi_{J}^{bj}(m,n,{\bf k}){\cal N}^{'}(n)e^{-qx^2/2}H_n(\sqrt{q}x)e^{-i(\o_m\tau+{\bf k.y})}\nonumber\\
&\times& N^{1/2}\int\f{d^2{\bf k}}{(2\pi\sqrt{q})^2}\sum_{m,n}\left[C^c(m,n,{\bf k})\phi_n(x)+
\tilde{A}^{2c}_1(m,n,{\bf k})\tilde{\phi}_n(x)\right]e^{-i(\o_m\tau+{\bf k.y})}\nonumber\\
&\times& N^{1/2}\int\f{d^2{\bf k}}{(2\pi\sqrt{q})^2}\sum_{m,n}\Phi_{J}^{dl}(m,n,{\bf k}){\cal N}^{'}(n)e^{-qx^2/2}H_n(\sqrt{q}x)e^{-i(\o_m\tau+{\bf k.y})}\nonumber\\
&\times& \tr\left\{\lambda^a,\lambda^b\right\}\left\{\lambda^c,\lambda^d\right\}\nonumber\\
&+& \hf
N^{1/2}\int\f{d^2{\bf k}}{(2\pi\sqrt{q})^2}\sum_{m,n}\left[C^a(m,n,{\bf k})\phi_n(x)+
\tilde{A}^{2a}_1(m,n,{\bf k})\tilde{\phi}_n(x)\right]e^{-i(\o_m\tau+{\bf k.y})}\nonumber\\
&\times& N^{1/2}\int\f{d^2{\bf k}}{(2\pi\sqrt{q})^2}\sum_{m,n}\Phi_{J}^{b1}(m,n,{\bf k}){\cal N}^{'}(n)e^{-qx^2/2}H_n(\sqrt{q}x)e^{-i(\o_m\tau+{\bf k.y})}\nonumber\\
&\times& N^{1/2}\int\f{d^2{\bf k}}{(2\pi\sqrt{q})^2}\sum_{m,n}\left[C^c(m,n,{\bf k})\phi_n(x)+
\tilde{A}^{2c}_1(m,n,{\bf k})\tilde{\phi}_n(x)\right]e^{-i(\o_m\tau+{\bf k.y})}\nonumber\\
&\times& N^{1/2}\int\f{d^2{\bf k}}{(2\pi\sqrt{q})^2}\sum_{m,n}\Phi_{J}^{d1}(m,n,{\bf k}){\cal N}^{'}(n)e^{-qx^2/2}H_n(\sqrt{q}x)e^{-i(\o_m\tau+{\bf k.y})}\nonumber\\
&\times&\tr[\lambda^a,\lambda^b][\lambda^c,\lambda^d]\nonumber\\
&+& \hf
N^{1/2}\int\f{d^2{\bf k}}{(2\pi\sqrt{q})^2}\sum_{m,n}\left[C^a(m,n,{\bf k})\phi_n(x)+
\tilde{A}^{2a}_1(m,n,{\bf k})\tilde{\phi}_n(x)\right]e^{-i(\o_m\tau+{\bf k.y})}\nonumber\\
&\times& N^{1/2}\int\f{d^2{\bf k}}{(2\pi\sqrt{q})^2}\sum_{m,n}\Phi_{J}^{b0}(m,n,{\bf k}){\cal N}^{'}(n)e^{-qx^2/2}H_n(\sqrt{q}x)e^{-i(\o_m\tau+{\bf k.y})}\nonumber\\
&\times& N^{1/2}\int\f{d^2{\bf k}}{(2\pi\sqrt{q})^2}\sum_{m,n}\left[C^c(m,n,{\bf k})\phi_n(x)+
\tilde{A}^{2c}_1(m,n,{\bf k})\tilde{\phi}_n(x)\right]e^{-i(\o_m\tau+{\bf k.y})}\nonumber\\
&\times& N^{1/2}\int\f{d^2{\bf k}}{(2\pi\sqrt{q})^2}\sum_{m,n}\Phi_{J}^{d0}(m,n,{\bf k}){\cal N}^{'}(n)e^{-qx^2/2}H_n(\sqrt{q}x)e^{-i(\o_m\tau+{\bf k.y})}\nonumber\\
&\times&\tr[\lambda^a,\lambda^b][\lambda^c,\lambda^d].
\eeqa
Contributions to the diagram (A) in Figure \ref{tachampboson4pt} come from the first term with $j=l=2$, and the second term of \ref{Sbterm1}, whereas the contributions to the diagram (B) in Figure \ref{tachampboson4pt} come from first term with $j=l=3$ and the third term, in equation (\ref{Sbterm1}).

In addition to these, another term in the action, $\tr[A_{\mu},\Phi_I]^2$, on doing a similar calculation, also contributes to both diagrams in Figure \ref{tachampboson4pt} with tachyons coming from the mode expansion of fields $A^{a2}_1$ in the multiplet $\zt^a(x)$ written in (\ref{zeta1}).

\begin{figure}[h]
\begin{center}
\begin{psfrags}
\psfrag{a11}[][]{\scalebox{0.8}{$\Phi_J^{b1}/\Phi_J^{b2}$}}
\psfrag{a12}[][]{\scalebox{0.8}{$\Phi_J^{d1}/\Phi_J^{d2}$}}
\psfrag{c1}[][]{\scalebox{0.8}{$C^a(m^{''},n^{''},{\bf k}^{''})$}}
\psfrag{c2}[][]{\scalebox{0.8}{$C^c(\tilde{m}^{''},\tilde{n}^{''},\tilde{{\bf k}}^{''})$}}
\psfrag{v2}[][]{}
\begin{subfigure}[b]{0.13\textwidth}
\includegraphics[width=\textwidth,angle=0]{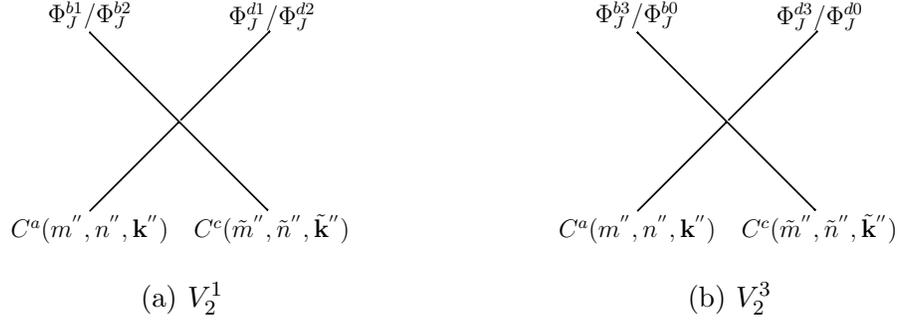}
\caption{$V^1_2$}
\label{3a}
\end{subfigure}
\end{psfrags}
~~~~~~~~~~~~~~~~~~~~~~~~~~~~~~~~
\begin{psfrags}
\psfrag{a11}[][]{\scalebox{0.8}{$\Phi_J^{b3}/\Phi_J^{b0}$}}
\psfrag{a12}[][]{\scalebox{0.8}{$\Phi_J^{d3}/\Phi_J^{d0}$}}
\psfrag{c1}[][]{\scalebox{0.8}{$C^a(m^{''},n^{''},{\bf k}^{''})$}}
\psfrag{c2}[][]{\scalebox{0.8}{$C^c(\tilde{m}^{''},\tilde{n}^{''},\tilde{{\bf k}}^{''})$}}
\psfrag{v2}[][]{}
\begin{subfigure}[b]{0.13\textwidth}
\includegraphics[width=\textwidth,angle=0]{vertex3.eps}
\caption{$V^{3}_2$}
\label{3b}
\end{subfigure}
\end{psfrags}
\caption{Four point vertices with two $C$ fields.}
\label{phiver}
\end{center}
\end{figure}

The interaction terms $ \hf \tr[\Phi_I,\Phi_J]^2 $ and $\tr[A_{\mu},\Phi_I]^2$ thus give the vertices as in Figure \ref{phiver}.
\\

The expression for $V^1_2$ as in Figure \ref{3a} is

\begin{eqnarray}
V^1_2&=&-\f{N}{2qg^2} (F^2_2+F^1_2)(n,n^{'},n^{''},\tilde{n}^{''}) (2\pi)^2\delta^2({\bf k}+{\bf k}^{'}+{\bf k}^{''}+\tilde{{\bf k}}^{''})\delta_{m+m^{'}+m^{''}+\tilde{m}^{''}}\times\non
&&\times(\tr[\lambda^a,\lambda^b][\lambda^c,\lambda^d]+\tr\{\lambda^a,\lambda^b\}\{\lambda^c,\lambda^d\})\non
&&F^1_2(n,n^{'},n^{''},\tilde{n}^{''})=\sqrt{q}({\cal N}^{'}(n))^2\int dx e^{-qx^2}\left [A_{n^{''}}(x)A_{\tilde{n}^{''}}(x)\right] \left[H_n(\sqrt{q}x)  H_{n^{'}}(\sqrt{q}x)\right]
\non
&&F^2_2(n,n^{'},n^{''},\tilde{n}^{''})=\sqrt{q}({\cal N}^{'}(n))^2\int dx e^{-qx^2}\left [\phi_{n^{''}}(x)\phi_{\tilde{n}^{''}}(x)\right] \left[H_n(\sqrt{q}x)  H_{n^{'}}(\sqrt{q}x)\right]
\end{eqnarray}

The expression for $V^3_2$ as in Figure \ref{3b} is

\begin{eqnarray}
V^{3}_2&=&-\f{N}{2qg^2} F^{3}_2(l,l^{'},n^{''},\tilde{n}^{''}) (2\pi)^2\delta^2({\bf k}+{\bf k}^{'}+{\bf k}^{''}+\tilde{{\bf k}}^{''})\delta_{m+m^{'}+m^{''}+\tilde{m}^{''}}\times\non
&&\times(\tr[\lambda^a,\lambda^b][\lambda^c,\lambda^d]+\tr\{\lambda^a,\lambda^b\}\{\lambda^c,\lambda^d\})\non
&&F^{3}_2(l,l^{'},n^{''},\tilde{n}^{''})=\sqrt{q}\int dx \left [A_{n^{''}}(x) A_{\tilde{n}^{''}}(x)+\phi_{n^{''}}(x)\phi_{\tilde{n}^{''}}(x)\right] \left[e^{-ilx}e^{-il^{'}x}\right]
\end{eqnarray}

%We also have additional vertices as $V^{3}_2$ for the $A_i^{3}$ $(i\ne1)$ fields in place of $\Phi_I^{3}/\tilde{\Phi_I}^{3}$ fields.

In the above expressions, $\mathcal{N}^{'}(n)=1/\sqrt{\sqrt{\pi}2^n n!}$. The functions $A_n(x)$ and $\phi_n(x)$ are defined in Appendix \ref{eigenfns}.

Further, the  propagators for the fields $\Phi_I^{a1,a2}$, $\Phi_I^{a3,a0}$ ($I=2,3$) are
\beqa
\label{propphia1I}
\expect{\Phi_I^{a1,a2}(m,n,{\bf k})\Phi_I^{b1,b2}(m^{'},n^{'},{\bf k}^{'})}=qg^2\f{\delta_{m,-m^{'}}\delta_{n,n^{'}}(2\pi)^2\delta^2({\bf k}+{\bf k}^{'})\delta^{ab}}{\o_m^2+\gamma_n+|{\bf k}|^2},
\eeqa
\beqa
\label{propphia3I}
\expect{\Phi_I^{a3,a0}(m,l,{\bf k})\Phi_I^{b3,b0}(m^{'},l^{'},{\bf k}^{'})}=qg^2 \f{\delta_{m,-m^{'}}2\pi\delta(l+l^{'})(2\pi)^2\delta^2({\bf k}+{\bf k}^{'})\delta^{ab}}{\o_m^2+l^2+|{\bf k}|^2}
\eeqa

Other than the terms considered so far, the term $[\Phi_I,\tilde{\Phi}_J]^2$ in action, (\ref{actionboson}), also contribute to the amplitudes in Figure \ref{tachampboson4pt}. The corresponding vertices and propagators for the fields $\tilde{\Phi}_J^{b1}/\tilde{\Phi}_J^{b2}$ and $\tilde{\Phi}_J^{b3}/\tilde{\Phi}_J^{b0}$ are same as those written above.

Using the vertices and propagator written above, and with $C^a$ and $C^c$ as external fields (and accordingly, $b=d$), the desired amplitudes come out to be
\beqa\label{tachampbosonI}
(A) = \hf MN\sum_{m,n} \int \f{d^2 {\bf k}}{(2\pi\sqrt{q})^2} (F_2^{2}+F_2^1)(0,0,n,n)\f{5}{\o_m^2 + \gamma_n+|{\bf k}|^2}\delta^{ac}
\eeqa
and
\beqa\label{tachampbosonII}
(B) = \hf MN\sum_{m}\int \f{d l}{(2\pi\sqrt{q})}\int \f{d^2 {\bf k}}{(2\pi\sqrt{q})^2}F_2^3(0,0,l,-l) \f{5}{(\o_m^2+l^2+|{\bf k}|^2)}\delta^{ac}.
\eeqa
The factor of $5$ appears because the flavor index $I$ in $\Phi^{a1,a2}_I$ runs over $2,3$ while that in $\tilde{\Phi}^{a1,a2}_I$ runs over $1,2,3$. We observe that substituting $M=1$ in above amplitude gives back the corresponding expressions, in (F.4) and (F.5) respectively of Appendix F.1, in \cite{1}.\\

\begin{figure}[h]
\begin{center}
\begin{psfrags}
\psfrag{c1}[][]{$C^a$}
\psfrag{c2}[][]{$C^c$}
\psfrag{c3}[][]{$\Phi_I^{b1}/\Phi_I^{b2}/\tilde{\Phi}^{b1}_I/\tilde{\Phi}^{b2}_I$}
\psfrag{c4}[][]{$\Phi_I^{b3}/\Phi_I^{b0}/\tilde{\Phi}_I^{b3}/\tilde{\Phi}_I^{b0}$}
\psfrag{v1}[][]{$V_5$}
\psfrag{v2}[][]{$V^{*}_5$}
\includegraphics[width= 8cm,angle=0]{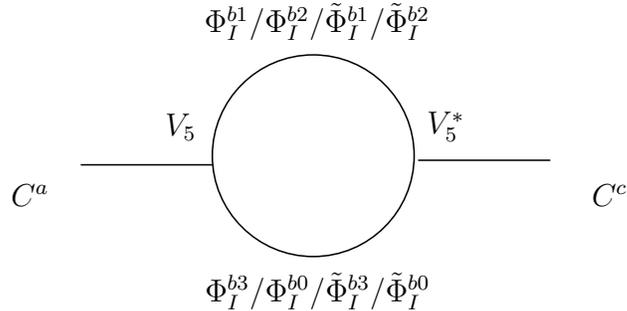}
\caption{A tachyon amplitude with bosonic three-point vertices}
\label{tachampboson3pt}
\end{psfrags}
\end{center}
\end{figure}

Next, we consider the tachyon two-point amplitude constructed out of bosonic three-point vertices in Figure \ref{tachampboson3pt}.

Consider the term $-2i\tr(\partial^{\mu}\Phi_I[A_{\mu},\Phi_I])$ in the action equation \ref{actionboson}. Writing the fields using $ SU(2M) $ generators,
\beqa
-2i\tr\left\{\partial^{\mu}\Phi^{ai}_I\f{1}{\sqrt{2}} (\lambda^a \otimes \sigma^i)\left[A_{\mu}^{bj}\f{1}{\sqrt{2}}(\lambda^b\otimes\sigma^j),\Phi^{ck}_I\f{1}{\sqrt{2}}(\lambda^c\otimes\sigma^k)\right]\right\}.
\eeqa
We make the choice $j=2$, $\mu=1$ for tachyon. Simplifying the commutator gives
\beqa
-2i\partial^{1}\Phi^{ai}_IA_1^{b2}\Phi^{ck}_I\tr\left\{\f{1}{\sqrt{2}} (\lambda^a \otimes \sigma^i)\left(\hf[\lambda^b,\lambda^c]\otimes[\sigma^2,\sigma^k]+\hf\lambda^c\lambda^b\otimes[\sigma^2,\sigma^k]+\hf[\lambda^b,\lambda^c]\otimes\sigma^k\sigma^2\right)\right\}.
\eeqa
Performing the sum over $i$ and $k$ and simplifying the commutators of Pauli matrices followed by tracing over them gives
\beqa
\hspace{-2mm}\sqrt{2}\left(\partial^1\Phi^{a1}_IA_1^{b2}\Phi^{c3}_I-\partial^1\Phi^{a3}_IA_1^{b2}\Phi^{c1}_I\right)\tr(\lambda^a\{\lambda^b,\lambda^c\})-i\sqrt{2}\left(\partial^1\Phi^{a2}_IA_1^{b2}\Phi^{c0}_I+\partial^1\Phi^{a0}_IA_1^{b2}\Phi^{c2}_I\right)\tr(\lambda^a[\lambda^b,\lambda^c]).
\eeqa

Above interaction terms contribute to the following vertex and propagators for $\Phi_I^{a1,a2}$ and $\Phi_I^{a3,a0}$ $(I=2,3)$ fields.
\begin{table}[H]
\begin{center}
\begin{tabular}{lcc}
\begin{psfrags}
\psfrag{c}[][]{$C^b(m^{''},n^{''},{\bf k}^{''})$}
\psfrag{a1}[][]{$\Phi_I^{a1}/\Phi_I^{a2}(m,n,{\bf k})$}
\psfrag{a3}[][]{$\Phi_I^{c3}/\Phi_I^{c0}(m^{'},l^{'},{\bf k}^{'})$}
\psfrag{v4}[][]{$V_5$}
\parbox[c]{5cm}{\includegraphics[width= 5 cm,angle=0]{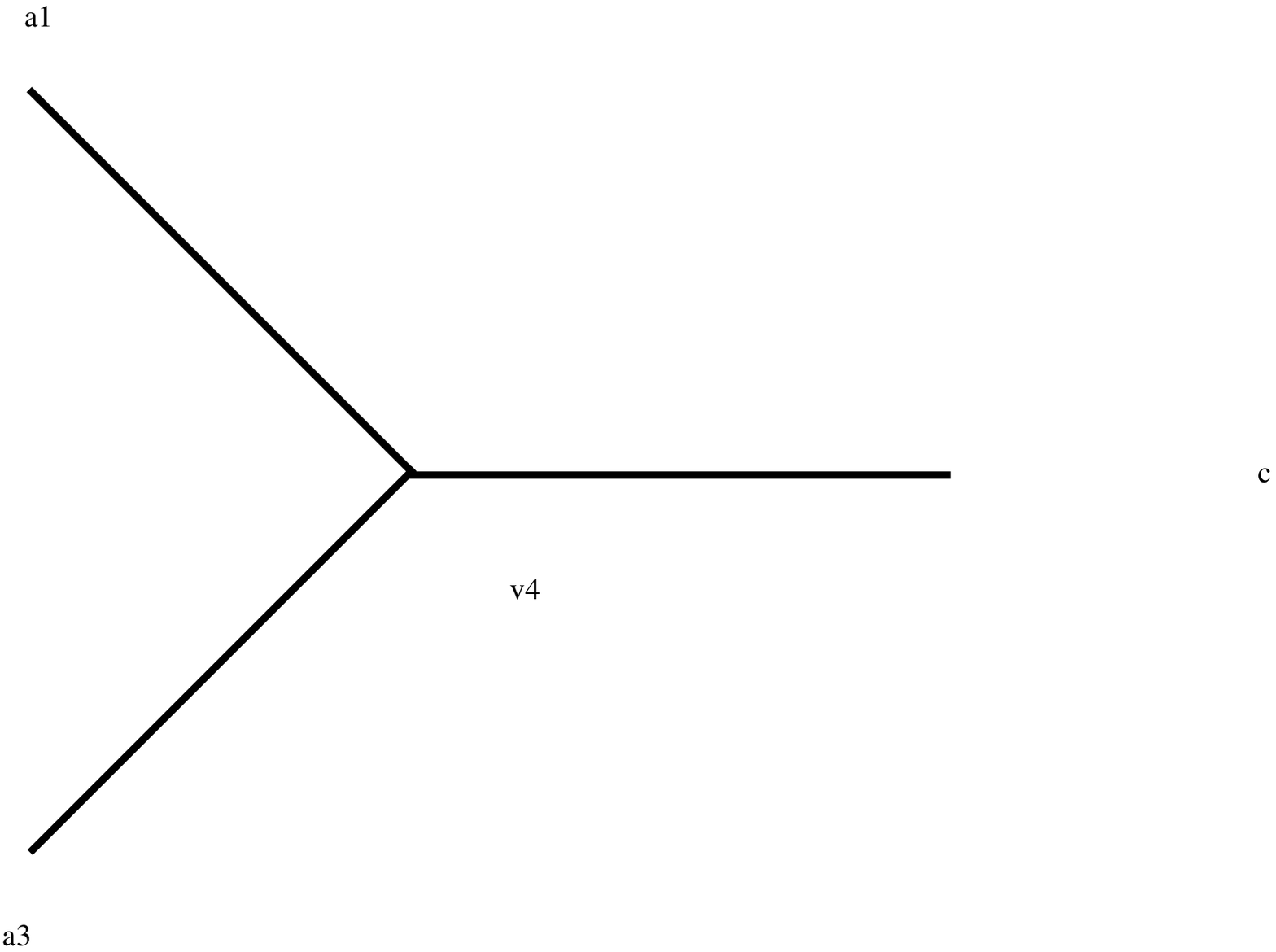}} 
\end{psfrags}
&~~~&
\parbox[c]{10cm}{$\begin{array}{c}V_5 =-\f{N^{3/2}}{qg^2} \beta F_5(n,l^{'},n^{''}) (2\pi)^2\delta^2({\bf k}+{\bf k}^{'}+{\bf k}^{''})\delta_{m+m^{'}+m^{''}}\\
\times\tr(\lambda^a[\lambda^b,\lambda^c]) \mbox{~~~ for $\Phi_I^{c0}$}\\
=-\f{N^{3/2}}{qg^2} \beta F_5(n,l^{'},n^{''}) (2\pi)^2\delta^2({\bf k}+{\bf k}^{'}+{\bf k}^{''})\delta_{m+m^{'}+m^{''}}\\
\times\tr(\lambda^a\{\lambda^b,\lambda^c\}) \mbox{~~~ for $\Phi_I^{c3}$}\\
F_5(n,l^{'},n^{''})=\int dx   \left[A_{n^{''}}(x)\partial_x H_n(\sqrt{q}x)-(qx)A_{n^{''}}(x) H_n(\sqrt{q}x)\right.\\
+ \left.(il^{'})A_{n^{''}}(x) H_n(\sqrt{q}x)
\right]e^{-qx^2/2}e^{-il^{'}x}{\cal N}^{'}(n),\end{array}$}
\end{tabular}
\end{center}
\end{table}

Similar expressions hold for vertex and propagator for $\tilde{\Phi}_I^{a1,a2}/\tilde{\Phi}_I^{a3,a0}$ fields. 
The relevant propagators are given in equations (\ref{propphia1I}) and (\ref{propphia3I}). Using the propagators and vertices written above, with tachyons $C^a$ and $C^c$ as external fields, the desired amplitude comes out to be

%\beqa
%\label{propphia1I}
%\expect{\Phi_I^{a1,a2}(m,n,{\bf k})\Phi_I^{b1,b2}(m^{'},n^{'},{\bf k}^{'})}=qg^2\f{\delta_{m,-m^{'}}\delta_{n,n^{'}}(2\pi)^2\delta^2({\bf k}+{\bf k}^%{'})\delta^{ab}}{\o_m^2+\gamma_n+|{\bf k}|^2}
%\eeqa

%\beqa
%\label{propphia3I}
%\expect{\Phi_I^{a3,a0}(m,l,{\bf k})\Phi_I^{b3,b0}(m^{'},l^{'},{\bf k}^{'})}=qg^2 \f{\delta_{m,-m^{'}}2\pi\delta(l+l^{'})(2\pi)^2\delta^2({\bf k}+{\bf %k}^{'})\delta^{ab}}{\o_m^2+l^2+|{\bf k}|^2}
%\eeqa

\beqa
\hspace{-1.5cm}-\hf M N \sum_{m,n}\int \f{d l}{(2\pi\sqrt{q})} \int \f{d^2 {\bf k}}{(2\pi\sqrt{q})^2}5\left[\left(q F_5(0,n,l)F_5(0,n,-l)\right)
\f{1}{(\o_m^2+\gamma_n+|{\bf k}|^2)(\o_m^2+l^2+|{\bf k}|^2)}\right]\delta^{ac}.
\eeqa
It can easily be seen that for $M=1$, above amplitude conforms with the corresponding expression, in (F.9) of Appendix F.1, in \cite{1}.\\

\begin{figure}[h]
\begin{center}
\begin{psfrags}
\psfrag{c2}[][]{\scalebox{0.85}{$C^a(\tilde{\o}_{m^{''}},\tilde{n}^{''},\tilde{{\bf k}}^{''})$}}
\psfrag{c1}[][]{\scalebox{0.85}{$C^c(\o_{m^{''}},n^{''},{\bf k}^{''})$}}
\psfrag{c3}[][]{\scalebox{0.85}{$\theta_1^b(m,n,{\bf k})$}}
\psfrag{c4}[][]{\scalebox{0.85}{$\lambda^{b3}_1(m^{'},n^{'},{\bf k}^{'})$}}
\psfrag{v1}[][]{\scalebox{0.85}{$F_6$}}
\psfrag{v2}[][]{\scalebox{0.85}{$F_6^{*}$}}
\psfrag{v3}[][]{\scalebox{0.85}{$V^{'\mu}_{f}$}}
\psfrag{v4}[][]{\scalebox{0.85}{$V^{'\mu*}_{f}$}}
\psfrag{a1}[][]{(a)}
\psfrag{a2}[][]{(b)}
\includegraphics[ width= 7cm,angle=0]{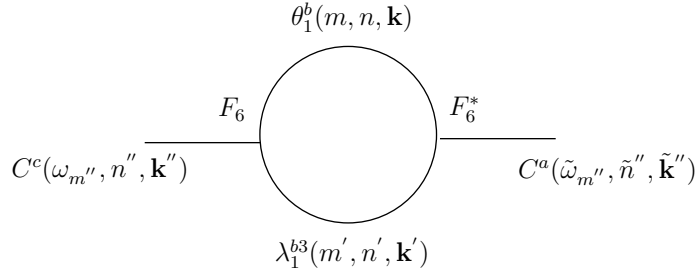}
\end{psfrags}
\caption{{Feynman diagrams involving three-point fermionic vertices}}
\label{tachampfermion}
\end{center}
\end{figure}

Now, we consider the tachyon two-point amplitude with fermions in the loop as shown in Figure \ref{tachampfermion}.

Consider the term $-i\tr\left(\bar{\lambda}_I\gamma^{\mu}[A_{\mu},\lambda_I]\right)$ in the fermionic action, equation (\ref{actionboson}). Writing the fields using $SU(2M)$ generators,
\beqa
\tr\left\{\f{1}{\sqrt{2}}(\lambda^a\otimes\sigma^i)\bar{\lambda}^{ai}_I\gamma^{\mu}\left[A^{bj}_{\mu}\f{1}{\sqrt{2}}(\lambda^b\otimes\sigma^j),\lambda^{ck}_I\f{1}{\sqrt{2}}(\lambda^c\otimes\sigma^k)\right]\right\}.
\eeqa
Further, note that tachyonic modes exist only for $j=2$, $\mu=1$. Simplifying the commutator gives
\beqa
\f{1}{2\sqrt{2}}\bar\lambda^{ai}_I\gamma^1A^{b2}_1\lambda^{ck}_I\tr\left\{(\lambda^a\otimes\sigma^i)([\lambda^b,\lambda^c]\otimes[\sigma^2,\sigma^k]+\lambda^c\lambda^b[\sigma^2,\sigma^k]+[\lambda^b,\lambda^c]\otimes\sigma^k\sigma^2)\right\}.
\eeqa

\begin{figure}[h]
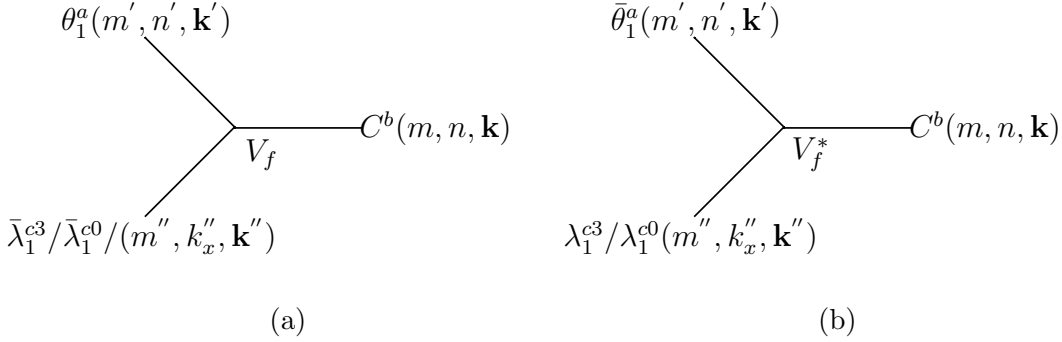

\begin{center}
\begin{psfrags}
\psfrag{c}[][]{$C^b(m,n,{\bf k})$}
\psfrag{a1}[][]{$\theta_1^a(m^{'},n^{'},{\bf k}^{'})$}
\psfrag{a3}[][]{$\bar{\lambda}^{c3}_1/\bar{\lambda}^{c0}_1/(m^{''},k_x^{''},{\bf k}^{''})$}
\psfrag{v4}[][]{$V_{f}$}
\begin{subfigure}[b]{0.2\textwidth}
\includegraphics[width=\textwidth,angle=0]{vertex1.eps} 
\caption{}
\label{5a}
\end{subfigure}
\end{psfrags}
~~~~~~~~~~~~~~~~~~~~~~
\begin{psfrags}
\psfrag{c}[][]{$C^b(m,n,{\bf k})$}
\psfrag{a1}[][]{$\bar{\theta}^a_1(m^{'},n^{'},{\bf k}^{'})$}
\psfrag{a3}[][]{$\lambda^{c3}_1/\lambda^{c0}_1(m^{''},k_x^{''},{\bf k}^{''})$}
\psfrag{v4}[][]{$V^{*}_{f}$}
\begin{subfigure}[b]{0.2\textwidth}
\includegraphics[width=\textwidth,angle=0]{vertex1.eps} 
\caption{}
\label{5b}
\end{subfigure}
\end{psfrags}
\caption{Fermion vertices}
\label{fermionv}
\end{center}
\end{figure}

A similar calculation holds for the term $\tr\left(\bar{\lambda}_K[\alpha^I_{KL}\Phi_I,\lambda_L]\right)$. These together contribute to the following two vertices and propagators for $\theta_i^a$ $(i=1,...,4)$ and $\lambda_l^{a3}$ $(l=1,...,4)$ fields as shown in Figure \ref{fermionv}.
\\

The expression for $V_f$ as in Figure \ref{5a} is,

\begin{eqnarray}
V_f &=&i\sqrt{2}\f{N}{qg^2}\gamma^{1} F_6(n,n^{'},k^{''}_x)(2\pi)^2\delta^2({\bf k}+{\bf k}^{'}-{\bf k}^{''})\delta_{m+m^{'}-m^{''}}\times\tr(\lambda^c\{\lambda^b,\lambda^a\})\mbox{~~for $\l^{c0}_1$}\non
&=&-\sqrt{2}\f{N}{qg^2}\gamma^{1} F_6(n,n^{'},k^{''}_x)(2\pi)^2\delta^2({\bf k}+{\bf k}^{'}-{\bf k}^{''})\delta_{m+m^{'}-m^{''}}\times\tr(\lambda^c[\lambda^b,\lambda^a])\mbox{~~for $\l^{c3}_1$}\non
&&\mbox{where~~} F_6(n,n^{'},k^{''}_x)=\sqrt{q}\int dx  \left(L_{n^{'}}(x)A_{n}(x)+R_{n^{'}}(x)\phi_{n}(x)\right)e^{ik^{''}_x x}.
\end{eqnarray}

Similarly $V^*_f$ as in Figure \ref{5b} is,

\begin{eqnarray}
V^{*}_f &=&-\sqrt{2}i\f{N}{qg^2}\gamma^{1} F^{*}_6(n,n^{'},k^{''}_x)(2\pi)^2\delta^2({\bf k}-{\bf k}^{'}+{\bf k}^{''})\delta_{m-m^{'}+m^{''}}\times\tr(\lambda^a\{\lambda^b,\lambda^c\})\mbox{~~for $\l^{c0}_1$}\non
&=&-\sqrt{2}\f{N}{qg^2}\gamma^{1} F^{*}_6(n,n^{'},k^{''}_x)(2\pi)^2\delta^2({\bf k}-{\bf k}^{'}+{\bf k}^{''})\delta_{m-m^{'}+m^{''}}\times\tr(\lambda^a[\lambda^b,\lambda^c])\mbox{~~for $\l^{c3}_1$}\non
&&\mbox{where~~} F^{*}_6(n,n^{'},k^{''}_x)=\sqrt{q}\int dx  \left(L^{*}_{n^{'}}(x)A_{n}(x)+R^{*}_{n^{'}}(x)\phi_{n}(x)\right)e^{-ik^{''}_x x}.
\end{eqnarray}

The functions $L_n(x), R_n(x), A_n(x)$ and $\phi_n(x)$ appearing in the above two vertices are defined in Appendix \ref{eigenfns}. The relevant propagators for the computation of one-loop amplitude are,

\beqa
\label{thetaai}
\expect{\theta_i^a(m,n,{\bf k})\theta_j^b(m^{'},n^{'},{\bf k}^{'})}=-iqg^2\f{\delta_{m,-m^{'}}\delta_{n,n^{'}}(2\pi)^2\delta^2({\bf k}+{\bf k}^{'})\delta^{ab}\delta_{ij}}{\slashed{P}_+}
\eeqa

\beqa
\label{lambdaa3i}
\expect{\lambda_i^{a3}(m,n,{\bf k})\lambda_j^{b3}(m^{'},n^{'},{\bf k}^{'})}=-iqg^2\f{\delta_{m,-m^{'}}\delta_{n,n^{'}}(2\pi)^2\delta^2({\bf k}+{\bf k}^{'})\delta^{ab}\delta_{ij}}{\slashed{K}_+}
\eeqa
where $\slashed{P}_+=i\o_m\gamma^0+\sqrt{\lambda^{'}_n}\gamma^1+k_2\gamma^2+k_3\gamma^3$ and $\slashed{K}_+=i\o_m\gamma^0+k_x\gamma^1+k_2\gamma^2+k_3\gamma^3$.

In addition to the above, there are three sets of similar vertices coming from other three multiplets written in equation (\ref{fmultiplets}) and corresponding propagators. Taking into account all these, with $C^a$ and $C^c$ as external fields, the desired one-loop amplitude comes out as

\beqa\label{tachfermion}
\Sigma^3_{C-C}=-4 MN \sum_{m,n}\int \frac{d^2 {\bf k}}{(2\pi\sqrt{q})^2}\frac{dl}{(2\pi\sqrt{q})}\mbox{tr}\left[F_6(n^{''},n,l)\gamma^1\frac{1}{\slashed{P}_{+}}F^{*}_6(\tilde{n}^{''},n,l)\gamma^1\frac{1}{\slashed{K}^{'}_{+}}
\right]\\ \nonumber
\times(2\pi)^2\delta^{2}({\bf k}^{''}+\tilde{{\bf k}}^{''})\delta_{m^{''}+\tilde{m}^{''}}\delta^{ac},
\eeqa
where trace is over the fermion indices. Again, note that the above expression for $M=1$ is in conformity with the corresponding result in (3.10) of \cite{1}.

It can thus be seen from all the sample computations that have been done, the full two point tachyon amplitude can be written as:

\beqa\label{cfactor1}
\Sigma^2_{tachyon}=\left(M\right) \times \Sigma^2
\eeqa

where $M$ is the color factor that arises from the unbroken gauge symmetry and $\Sigma^2$ is the one-loop two-point amplitude computed in \cite{1}.

\section{UV and IR divergences}\label{uvir}

We have seen in the previous section that the two-point amplitude for the tachyons is equal to a color factor times the amplitude obtained in \cite{1}. The color factor comes from the unbroken gauge group. Thus

\beqa\label{cfactor}
\Sigma^2_{tachyon}=\left(\mbox{color factor}\right) \times \Sigma^2
\eeqa

where $\Sigma^2$ is the amplitude obtained in \cite{1}. The individual loop diagrams that have been computed in the previous section are divergent both in the UV as well as in the IR. Let us now pause to discuss on how to address these divergences.

The ${\cal N}=4$ supersymmetry is broken completely by the background expectation value of the scalar. Due to the finiteness of the background value of the scalar, the ultraviolet properties of the ${\cal N}=4$
theory are not affected by this breaking of supersymmetry. Since the ${\cal N}=4$ theory is finite, the broken theory also is UV finite. This can be checked by looking at the ultraviolet behavior of the full amplitude with both bosons and fermions in the loop. The UV contributions from bosons and fermions cancel. This cancellation at the one-loop level was demonstrated in detail in \cite{1} and \cite{2}. This computation carries over trivially to the present case.

The IR divergences appear from two sources. One is the artifact of the temporal, $A_0=0$ gauge. These need to be removed following the prescription as discussed in \cite{kapusta1}-\cite{Leib1}. The others are genuine IR divergences appearing from the massless fields propagating in the loop. These are the fields that do not couple to the background scalar at the quadratic level. In the $SU(2M) \rightarrow SU(M)\times SU(M)\times U(1)$ case, since the background scalar is $\Phi_1=qx \f{1}{\sqrt{2}}\left(\l^0\otimes\s^3\right)$, the tree-level massless fields are $\Phi_I^{a0}/\tilde{\Phi}_I^{a0}$, $\Phi_I^{a3}/\tilde{\Phi}_I^{a3}$ $(I=1,2,3)$, $A_{i}^{a0}$ and $A_{i}^{a3}$ $(i=1,2,3)$. In \cite{1} and \cite{2} these IR divergences were dealt with in two steps.
We first computed the one-loop corrected propagator for these massless fields at finite temperature. This propagator in turn gives corrected masses for the tree-level massless fields. The corrected propagator is then used to compute the tachyon two-point amplitude. The resulting amplitude is thus IR finite. An apparent additional complication in the present case is that the number of massless fields are more. However since these fields transform as adjoint of $SU(M)$, the unbroken symmetry, the number of one-loop diagrams that need to be computed is essentially the same as in \cite{1}. The only difference here being the appearance of the extra color factors as in (\ref{cfactor}). We however do not repeat this exercise here. This exact computation will be necessary for calculating the transition temperature at which the tachyon becomes massless and the stacks of branes become stable. Presently we wish to settle with a more modest goal of establishing as to how our previous computations can be adapted to the present case of intersecting stacks.

In the following sections we shall concentrate on only computation of the color factor factors that arise in more general setups. The issues related to UV and IR divergences in these cases can easily be addressed as discussed above.

\section{$SU(M_1+M_2)\rightarrow SU(M_1)\times SU(M_2)\times U(1)$}\label{M1M2} 

In this section we generalize the computations of the previous sections to the case of $SU(M_1+M_2)\rightarrow SU(M_1)\times SU(M_2)\times U(1)$ breaking. To proceed we first write down the generators of $SU(M_1+M_2)$. Let us denote the generators of $SU(M_1)$ by $\l_1^{\bm{a}_1}$ with 
$a_1=1, \cdots M_1^2-1$ and those of  $SU(M_2)$ by $\l_2^{\bm{a}_2}$ with $a_{2}=1, \cdots M_2^2-1$. These generators satisfy $\tr(\l_1^{\bm{a}_1}\l_1^{\bm{b}_1})=\f{1}{2}\d^{\bm{a}_1\bm{b}_1}$ and $\tr(\l_2^{\bm{a}_2}\l_2^{\bm{b}_2})=\f{1}{2}\d^{\bm{a}_2\bm{b}_2}$. 
The background scalar will be chosen to be $\Phi_1^D=qx T_D$, with

\begin{eqnarray}
T_D=\scalemath{0.7}{\sqrt{\frac{1}{2(M_1+M_2)}}}\left(\scalemath{0.7}{
\begin{BMAT}[8pt]{cccccc}{cccccc}
\sqrt{\frac{M_2}{M_1}}&0&\cdots&0&0&\cdots\\
0&\sqrt{\frac{M_2}{M_1}}&\cdots&0&0&\cdots\\
\vdots&\vdots&\ddots&\vdots&\vdots&\ddots\\
0&0&\cdots&-\sqrt{\frac{M_1}{M_2}}&0&\cdots\\
0&0&\cdots&0&-\sqrt{\frac{M_1}{M_2}}&\cdots\\
\vdots&\vdots&\ddots&\vdots&\vdots&\ddots
\end{BMAT}}\right)
\end{eqnarray}

The other unbroken generators of $SU(M_1+M_2)$ are then
\begin{eqnarray}
T_1^{a_1}=\left(\scalemath{0.7}{
\begin{BMAT}[8pt]{ccc:ccc}{ccc:ccc}
&&&0&0&\cdots\\
&\makebox[\wd0]{\large $\l_1^{a_1}$}&&0&0&\cdots\\
&&&\vdots&\vdots&\ddots\\
0&0&\cdots&&&\\
0&0&\cdots&&\makebox[\wd0]{\large $0_{M_2\times M_2}$}&\\
\vdots&\vdots&\ddots&&&\\
\end{BMAT}}\right)~~~;~~~
T_2^{a_2}=\left(\scalemath{0.7}{
\begin{BMAT}[8pt]{ccc:ccc}{ccc:ccc}
&&&0&0&\cdots\\
&\makebox[\wd0]{\large $0_{M_1\times M_1}$}&&0&0&\cdots\\
&&&\vdots&\vdots&\ddots\\
0&0&\cdots&&&\\
0&0&\cdots&&\makebox[\wd0]{\large $\l_2^{a_2}$}&\\
\vdots&\vdots&\ddots&&&\\
\end{BMAT}}\right)
\end{eqnarray}

The $2M_1M_2$ broken generators will be denoted by $T_3^{\a_1}$, $(\a_1=1,\dots,2M_1M_2)$. These broken generators can be grouped into pairs. The explicit form of a pair is

\begin{eqnarray}
\label{offD}
T_3^1=\frac{1}{2}\left(\scalemath{0.7}{
\begin{BMAT}[8pt]{ccc:ccc}{ccc:ccc}
&&&1&0&\cdots\\
&\makebox[\wd0]{\large $0_{M_1\times M_1}$}&&0&0&\cdots\\
&&&\vdots&\vdots&\ddots\\
1&0&\cdots&&&\\
0&0&\cdots&&\makebox[\wd0]{\large $0_{M_2\times M_2}$}&\\
\vdots&\vdots&\ddots&&&\\
\end{BMAT}}\right)
~~~;~~~T_3^2=\frac{1}{2}\left(
\scalemath{0.7}{\begin{BMAT}[8pt]{ccc:ccc}{ccc:ccc}
&&&-i&0&\cdots\\
&\makebox[\wd0]{\large $0_{M_1\times M_1}$}&&0&0&\cdots\\
&&&\vdots&\vdots&\ddots\\
i&0&\cdots&&&\\
0&0&\cdots&&\makebox[\wd0]{\large $0_{M_2\times M_2}$}&\\
\vdots&\vdots&\ddots&&&\\
\end{BMAT}}\right)
\end{eqnarray}

Likewise there are $M_1M_2$ pairs with $1$ or $\pm i$ placed in element of the off-diagonal blocks. The adjoint fields are now written as

\beqa
A_{\mu}=A_{\mu}^uT^u=A_{\mu}^{D}T_D+A_{\mu}^{1a_1}T_1^{a_1}+A_{\mu}^{2a_2}T_2^{a_2}+A_{\mu}^{3\a_1}T_3^{\a_1},\non
\Phi_{I}=\Phi_{I}^uT^u=\Phi_{I}^{D}T_D+\Phi_{I}^{1a_1}T_1^{a_1}+\Phi_{I}^{2a_2}T_2^{a_2}+\Phi_{I}^{3\a_1}T_3^{\a_1},
\eeqa
where $u = 1,2,...,(M_1+M_2)^2-1$.

The computation of tree-level spectrum here, in the Yang-Mills approximation, is similar to what has been done in section \ref{treenn}. 
The $2M_1M_2$  tachyons corresponding to the broken generators occur as doublets of $(A,\Phi)$ fields as in equation (\ref{multiplet}). To see which 
ones form the pair let us look at the quadratic term (\ref{quadterm}). Expanding about the background scalar the quadratic terms for
the fluctuations are,
\beqa\label{expand2}
&-&2qi ~\tr\left(x\pa^{\mu}\Phi_1\left[A_{\mu},T_D\right]+T_D\left[A_1,\Phi_1\right]\right).
\eeqa

Noting that 
\beqa
[T_D,T_3^1]=i\sqrt{\f{M_1+M_2}{2M_1M_2}}T_3^2~~~ \mbox{and}~~~ [T_D,T_3^2]=-i\sqrt{\f{M_1+M_2}{2M_1M_2}}T_3^1 
\eeqa
and similarly for the other pairs of $T_3$ matrices, we can write (\ref{expand2}) as
\beqa\label{expand3}
&=& q\sqrt{\f{M_1+M_2}{2M_1M_2}}\left(x\pa^{\mu}\Phi_1^{31}A_{\mu}^{32}-x\pa^{\mu}\Phi_1^{32}A_{\mu}^{31}-\Phi_1^{31}A_1^{32}+\Phi_1^{32}A_1^{31}\right)+\cdots
\eeqa
The $\cdots$ in (\ref{expand3}) correspond to the terms involving the other pairs. As in section \ref{spectrum} we shall in the following absorb the factor of $\sqrt{\f{M_1+M_2}{2M_1M_2}}$ in equation (\ref{expand3}) into a redefinition of $q$.

Defining

\beqa\label{multiplet1}
\xi^{1}=\left(\begin{array}{c}\Phi^{31}_1\\A^{32}_1\\A^{32}_2\\A^{32}_3\end{array}\right)~;~\zeta^{1}=\left(\begin{array}{c}\Phi^{31}_1\\A^{32}_1\end{array}\right)~~~;~~~\xi^{'1}=\left(\begin{array}{c}\Phi^{32}_1\\A^{31}_1\\A^{31}_2\\A^{31}_3\end{array}\right)~;~\zeta^{'1}=\left(\begin{array}{c}\Phi^{32}_1\\A^{31}_1\end{array}\right)
\eeqa
(and similarly for the other $M_1M_2$ pairs) the quadratic action for bosons is written as
\beqa\label{actionb1}
S_b=\int d^4 z ~\left[\sum_{s=1}^{M_1M_2}\left(\f{1}{2}(\xi^{s})^T{\cal O}_B\xi^{s}+\f{1}{2}(\xi^{'s})^{T}{\cal O}^{'}_B\xi^{'s}\right)+{\cal L}(A^{D}_{\mu},A^{1a_1}_{\mu},
A^{2a_2}_{\mu},\Phi_I,\tilde{\Phi}_J)\right],
\eeqa
where ${\cal O}_B$ is defined in (\ref{ob}). Following section \ref{treenn} we have $2M_1M_2$ tachyons of mass-squared equal to $-q$. Apart from these tachyons there are other off-diagonal $A_{\mu}$ and $\Phi_I$ fields which are massive each with degeneracy $2M_1M_2$. These fields bear gauge components corresponding to those of the matrices (\ref{offD}) and are included in ${\cal L}(A^{D}_{\mu},A^{1a_1}_{\mu}, A^{2a_2}_{\mu},\Phi_I,\tilde{\Phi}_J)$ of (\ref{actionb1}). All the massive modes including the tachyons transform as bi-fundamentals under $SU(M_1)\times SU(M_2)$. The situation is analogous to the one represented in Figure \ref{intfig1}. These modes correspond to strings stretching between the two stacks of $M_1$ and $M_2$ branes. The massless fields correspond to the open strings with end points on the same stack. They transform as adjoint under $SU(M_1)$ or $SU(M_2)$ and under $U(1)$ generated by $T_D$. 

The analysis of the quadratic action for fermions also proceeds along the same lines as section \ref{treennf} with the modifications outlined as above.

In the following we now turn to the computation of a typical contribution to the one-loop tachyon two-point amplitudes. 
The main idea is to compute the color factors corresponding to the unbroken $SU(M_1)\times SU(M_2)\times U(1)$ symmetry. The full amplitude is then this color factor times the result obtained in \cite{1}, similar to what we have obtained in section \ref{amplitude}. We illustrate the computation here for scalars propagating in the loop. The calculation for gauge and fermions in the loop is analogous.

\begin{figure}[h]
\begin{center}
\begin{psfrags}
\psfrag{c1}[][]{$C$}
\psfrag{c2}[][]{$C$}
\psfrag{c4}[][]{$\Phi^{3\a_1}_I,\tilde{\Phi}^{3\a_1}_I$}
\psfrag{v2}[][]{$V^{2}_2$}
\includegraphics[width= 4cm,angle=0]{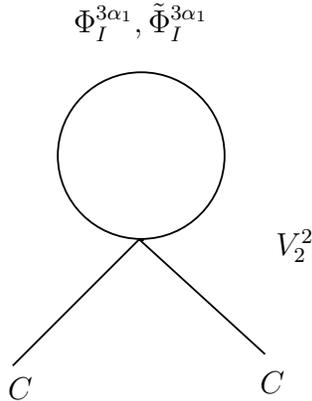}
\end{psfrags}
\caption{A tachyon two-point amplitude with four-point bosonic vertices}
\label{tachampbos4ptM1M2}
\end{center}
\end{figure}

Consider the diagram shown in Figure \ref{tachampbos4ptM1M2}. Where the $C$ fields are one of the $2M_1M_2$ the tachyons. 

For this, we again consider the term $ \hf \tr\left([\Phi_I,\Phi_J]^2\right) $ in the action, equation (\ref{actionboson}). Writing the fields using $ SU(M_1+M_2) $ generators,

\beqa
\hf \tr\left[\Phi^u_IT^u,\Phi^v_J T^v\right]\left[\Phi^w_I T^w,\Phi^s_J T^s\right]. \nonumber
\eeqa
As in section \ref{amplitude}, we make the choice $u=w=1$ and $I=1$ for tachyons on external legs. This amounts to the generator $T^1_3$ corresponding to one of the tachyons. Since there are two ways of choosing tachyons, we get
\beqa
(\Phi^{31}_1)^2\tr\left(\left[T^1_3,\Phi_J^DT_D+\Phi^{1a_1}_JT^{a_1}_1+\Phi^{2a_2}_JT^{a_2}_2+\Phi^{3\a_1}_JT^{\a_1}_3\right]^2\right).
\eeqa

Contribution to the required amplitude comes only from
\beqa\label{vertex4}
(\Phi^{31}_1)^2\Phi^{3\a_1}_J\Phi^{3\b_1}_J\tr\left([T^1_3,T^{\a_1}_3][T^1_3,T^{\b_1}_3]\right).
\eeqa
This contributes to the following vertices and propagators,

\begin{table}[H]
\begin{center}
\begin{tabular}{lcc}
\begin{psfrags}
\psfrag{a11}[][]{\scalebox{0.8}{$\Phi_J^{3\a_1}$}}
\psfrag{a12}[][]{\scalebox{0.8}{$\Phi_J^{3\b_1}$}}
\psfrag{c1}[][]{\scalebox{0.8}{$C(m^{''},n^{''},{\bf k}^{''})$}}
\psfrag{c2}[][]{\scalebox{0.8}{$C(\tilde{m}^{''},\tilde{n}^{''},\tilde{{\bf k}}^{''})$}}
\psfrag{v2}[][]{$V^2_2$}
\parbox[c]{3cm}{\includegraphics[width= 2.5cm,angle=0]{vertex3.eps}}
%\caption{$V_1$ , $V_2$ vertices}
\label{v1}
\end{psfrags}
&~~~&
\parbox[c]{12cm}{$\begin{array}{c}V^2_2=-\f{N}{qg^2} F^2_2(n,n^{'},n^{''},\tilde{n}^{''}) (2\pi)^2\delta^2({\bf k}+{\bf k}^{'}+{\bf k}^{''}+\tilde{{\bf k}}^{''})\delta_{m+m^{'}+m^{''}+\tilde{m}^{''}}\\
\times \tr[T^3_1,T^{\a_1}_3][T^3_1,T^{\b_1}_3]\\
%\times(\tr[\lambda^a,\lambda^b][\lambda^c,\lambda^d]+\tr\{\lambda^a,\lambda^b\}\{\lambda^c,\lambda^d\})\\
F^2_2(n,n^{'},n^{''},\tilde{n}^{''})=\sqrt{q}({\cal N}^{'}(n))^2\int dx e^{-qx^2}\left [\phi_{n^{''}}(x)\phi_{\tilde{n}^{''}}(x)\right] \left[H_n(x)  H_{n^{'}}(x)\right]\\
\\
\end{array}$}
\end{tabular}
\end{center}
\end{table}
and
\beqa
\label{propphialpha3I}
\expect{\Phi_I^{3\a_1}(m,n,{\bf k})\Phi_I^{3\b_1}(m^{'},n^{'},{\bf k}^{'})}=qg^2\f{\delta_{m,-m^{'}}\delta_{n,n^{'}}(2\pi)^2\delta^2({\bf k}+{\bf k}^{'})\delta^{\a_1\b_1}}{\o_m^2+\gamma_n+|{\bf k}|^2}.
\eeqa
The corresponding vertex and propagator for $\tilde{\Phi}_J^{3\alpha1}$ fields are same as those written above.

Now,
\beqa\label{factorinampM1M2}
\tr[T^1_3,T^{\a_1}_3][T^1_3,T^{\a_1}_3].
\eeqa
yields non-zero terms when $\a_1$ is such that the $M_1 \times M_2$ blocks in the off-diagonal generators $T^{\a_1}_3$ have non-zero elements in either first row or first column. We find that, for $\a_1=2$, \ref{factorinampM1M2} evaluates to $-\hf$, whereas for all other values of $\a_1$, it evaluates to $0$ or $-\f{1}{8}$. The number of elements in first row and first column together, of $M_1 \times M_2$ block, excluding the first element in the block, is $(M_1-1)+(M_2-1)$. Hence, $\ref{factorinampM1M2}$ evaluates to
$$-\hf - \f{1}{8}(M_1-1+M_2-1)\times 2 = -\f{M_1+M_2}{4}.$$
Using this, along with vertex and propagator, the required amplitude can be written as
\beqa\label{ampM1M2}
\left(\f{M_1+M_2}{2}\right)\hf N\sum_{m,n} \int \f{d^2 {\bf k}}{(2\pi\sqrt{q})^2} F_2^{2}(0,0,n,n)\f{5}{\o_m^2 + \gamma_n+|{\bf k}|^2}.
\eeqa

We now do the similar counting for the massless modes propagating in the loop corresponding to the diagram shown in Figure \ref{tachampbos4ptM1M2}.
The vertex has the form

\beqa\label{massless4}
(\Phi^{31}_1)^2\Phi^{3\a_1}_J\Phi^{3\b_1}_J\tr\left([T^1_3,T^{a}][T^1_3,T^{b}]\right).
\eeqa

The massless fields do not couple to the background. Thus, the generators $T^a$ and $T^b$ in the expression (\ref{massless4}) has non-zero 
entries only in the diagonal $M_1\times M_1$ and $M_2\times M_2$ blocks. With the tachyon corresponding to $T_3^1$, the generators with only
non zero entries only in the crossed places (that is the first rows in the $M_1\times M_1$ and $M_2\times M_2$ blocks) in the following matrix give finite contributions.

\begin{eqnarray}
\label{nonzeromassless}
\frac{1}{2}\left(\scalemath{0.7}{
\begin{BMAT}[8pt]{ccc:ccc}{ccc:ccc}
\times&\times&\cdots&0&0&\cdots\\
0&0&\cdots&0&0&\cdots\\
\vdots&\vdots&\ddots&\vdots&\vdots&\ddots\\
0&0&\cdots&\times&\times&\cdots\\
0&0&\cdots&0&0&\cdots\\
\vdots&\vdots&\ddots&\vdots&\vdots&\ddots\\
\end{BMAT}}\right)
\end{eqnarray}

The counting of these contributions is done as follows. For the $SU(M_!+M_2)$ there are $M_1+M_2-1$ diagonal (Cartan) generators. These generators have the form

\begin{eqnarray}\label{cartan1}
   H_{r}=\frac{1}{\sqrt{2r(r+1)}}\begin{tikzpicture}[decoration={brace,amplitude=5pt},baseline=(current bounding box.west)]
     \matrix (magic) [matrix of math nodes,left delimiter=(,right delimiter=)] {
      1 \\
      & 1 \\
      & & \ddots \\
      & & & 1 \\
      & & & & -r \\
      & & & & & 0 \\
      & & & & & & \ddots \\
      & & & & & & & 0\\
     };
     \draw[decorate] (magic-1-1.north) -- (magic-4-4.north east) node[above=5pt,midway,sloped] {$r$ times};
     
   \end{tikzpicture}
\end{eqnarray}

The contribution to the trace of the commutator in the vertex from $H_r$ is $-\frac{1}{4r(r+1)}$ for $r<M_1$. The sum of these contributions is

\begin{eqnarray}\label{cartan2}
-\frac{1}{4}\left[\frac{1}{1\cdot 2}+ \frac{1}{2\cdot 3}+\cdots+\frac{1}{M_1\cdot (M_1-1)}\right]=-\frac{1}{4}\left[1-\frac{1}{M_1}\right]
\end{eqnarray}
 
Similarly the contribution to the trace of the commutators from $H_{M_1}$ is 

\beqa\label{cartan3}
-\frac{1}{4}\left[1+\frac{1}{M_1}\right]. 
\eeqa

The other diagonal matrices give zero trace for the commutators.

We can now count the contributions from the other non-diagonal elements in (\ref{nonzeromassless}). This gives, 

\[-\left[\frac{M_1-1}{4}\right]-\left[\frac{M_2-1}{4}\right]\]

Thus all these sum up to

\[-\frac{1}{4}\left[1-\frac{1}{M_1}\right]-\frac{1}{4}\left[1+\frac{1}{M_1}\right]-\left[\frac{M_1-1}{4}\right]-\left[\frac{M_2-1}{4}\right]=-\frac{M_1+M_2}{4}\].

Now putting in the propagators, the amplitude with massless modes in the loop is

\beqa\label{ampM1M2massless}
\left(\f{M_1+M_2}{2}\right)\hf N\sum_{m}\int \f{d l}{(2\pi\sqrt{q})}\int \f{d^2 {\bf k}}{(2\pi\sqrt{q})^2}F_2^3(0,0,l,-l) \f{5}{(\o_m^2+l^2+|{\bf k}|^2)}.
\eeqa

A similar computation of the color factor can be done for the diagrams involving three-point vertices. To illustrate this computation let us consider
the following term 

\beqa\label{ver3ptm1m2}
-2i \tr \left(\partial^{\mu}\Phi_I\left[A_{\mu},\Phi_I\right]\right). 
\eeqa

\begin{figure}[h]
\begin{center}
\begin{psfrags}
\psfrag{m1}[][]{\scalebox{0.8}{\rotatebox{-45}{Massless}}}
\psfrag{m2}[][]{\scalebox{0.8}{\rotatebox{45}{Massive}}}
\psfrag{T}[][]{\scalebox{0.8}{\rotatebox{0}{Tachyon}}}
\includegraphics[width=4cm,angle=0]{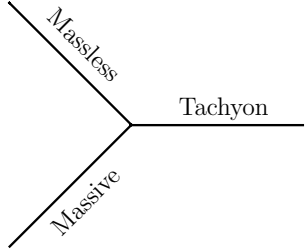}
\end{psfrags}
\caption{General structure of a three point vertex}
\label{ver3ptm1m2fig}
\end{center}
\end{figure}

Let us fix the gauge index for the $A_{\mu}$ field to be that of the generator $T_3^2$ (eqn. (\ref{offD})) so that this corresponds to the same tachyon mode as considered in the earlier parts of the section. 
Further let us take the $\Phi_I$ $(I\ne 1)$ field inside the commutator to be one of the massive modes. The vertex corresponding to (\ref{ver3ptm1m2}) is always of the form as shown in Figure \ref{ver3ptm1m2fig}.

We have seen that a non-zero commutator in (\ref{ver3ptm1m2}) results from generator corresponding to the $\Phi_I$ field being such that it has non-zero entries
in the first rows of the $M_1 \times M_2$ off-diagonal blocks. Excluding the first diagonal element in this off-diagonal block, the trace over these color indices for each of the generators corresponding to $\Phi_I$ in the commutator give a factor of $\frac{1}{2}$. In the one loop amplitude, the color factor from these generators is 

\beqa 
\frac{1}{4}\left[\left(M_1-1\right)+\left(M_2-1\right)\right].
\eeqa

The generator with first diagonal element in the $M_1 \times M_2$ off-diagonal block being non-zero is $T^1_3$ in (\ref{offD}). The result of the commutator of this generator with $T_3^2$ has a non-vanishing product with the Cartan generators. The counting is same as the one done for the case of four point vertices (see eqns (\ref{cartan1})-(\ref{cartan3})). The contribution from this is 

\beqa
\frac{1}{4}\left[1-\frac{1}{M_1}\right] +\frac{1}{4}\left[1+\frac{1}{M_1}\right].
\eeqa

The total color factor is then  $\frac{M_1+M_2}{4}$ as computed earlier. The rest of the computation is similar to that of \cite{1}. The answer for the one-loop two-point amplitude corresponding to the vertex (\ref{ver3ptm1m2}) with scalars in the loop is

\beqa
 -\left(\f{M_1+M_2}{2}\right)\hf N \sum_{m,n}\int \f{d l}{(2\pi\sqrt{q})} \int \f{d^2 {\bf k}}{(2\pi\sqrt{q})^2}(5)\times\left[
\f{\left(q F_5(0,n,l)F_5(0,n,-l)\right)}{(\o_m^2+\gamma_n+|{\bf k}|^2)(\o_m^2+l^2+|{\bf k}|^2)}\right].
\eeqa

We can repeat this computation for the other contributions coming from the diagrams with other fields in the loop. This however is not necessary for the present purpose as the computation of the color factor for the tachyon two-point amplitude is same as that for all the other modes (namely the gauge and the fermions) propagating in the loop. We thus conclude that 

\beqa\label{cfactor2}
\Sigma^2_{tachyon}= \left(\f{M_1+M_2}{2}\right)\times \Sigma^2,
\eeqa

where $\left(\f{M_1+M_2}{2}\right)$ is the color factor due to the unbroken gauge symmetry and $\Sigma^2$ is computed in \cite{1}.
This reduces to equation \ref{cfactor1} in section \ref{amplitude} on substituting $M_1=M_2=M$.

\section{Multiple Stacks}\label{multiple}

In this section, we extend the computation of the previous section to the case of $SU(M_1+M_2+... +M_p)\rightarrow SU(M_1)\times SU(M_2)\times...\times SU(M_p)\times U(1)^{p-1}$ symmetry breaking. We denote the generators of $SU(M_r)$ by $\l_r^{\bm{a}_r}$ with $\bm a_r=1, \cdots M_r^2-1$. Here, $r = 1, 2,... ,p$. As before, these generators satisfy $\tr(\l_r^{\bm{a}_r}\l_r^{\bm{b}_r})=\f{1}{2}\d^{\bm{a}_r\bm{b}_r}$.

We write the block-diagonal generators as
\[
T_r=\left(\scalemath{0.7}{
\begin{BMAT}[10pt]{ccccccc}{ccccccc}
(0)_{M_1\times M_1} & (0)_{M_1\times M_2} & \cdots & (0)_{M_1\times M_{r-1}} & (0)_{M_1\times M_r} & (0)_{M_1\times M_{r+1}} & \cdots\\
(0)_{M_1\times M_1} & (0)_{M_2\times M_2} & \cdots & (0)_{M_2\times M_{r-1}} & (0)_{M_1\times M_r} & (0)_{M_1\times M_{r+1}} & \cdots\\
\vdots & \vdots & \ddots & \vdots & \vdots & \vdots & \ddots\\
0 & 0 & \cdots & 0 & 0 & 0 & \cdots\\
0 & 0 & \cdots & 0 & (\lambda_r)_{M_r\times M_r} & 0 & \cdots\\
0 & 0 & \cdots & 0 & 0 & 0 & \cdots\\
\vdots & \vdots & \ddots & \vdots & \vdots & \vdots & \ddots
\end{BMAT}}\right).
\]
In addition, there are $p-1$ generators which are diagonal. The background field can be written using such a generator as $\Phi^D_1=qxT_D$, where $T_D$ is of the form

\[
\scalemath{0.7}~\left(\scalemath{0.7}{
\begin{BMAT}[8pt]{ccccccc}{ccccccc}
c_1&0&\cdots&0&\cdots&0&\cdots\\
\vdots&\ddots&\ddots&\vdots&\ddots&\vdots&\ddots\\
0&0&c_1&0&\cdots&0&\cdots\\
0&\cdots&0&c_2&0&0&\cdots\\
\vdots&\ddots&\vdots&\ddots&\ddots&\vdots&\ddots\\
0&\cdots&0&\cdots&0&c_2&\cdots\\
\vdots&\ddots&\vdots&\ddots&\vdots&\ddots&\ddots\\
\end{BMAT}}\right).
\]

\begin{figure}[t]
\begin{center}
\begin{psfrags}
\psfrag{1}[][]{$M_1$}
\psfrag{2}[][]{$M_2$}
\psfrag{p}[][]{$M_p$}
%\psfrag{A3}[][]{$A^{(a0,a3)}/\Phi_I^{(a0,a3)}$}
%\psfrag{A2}[][]{$A^{(a1,a2)}/\Phi_I^{(a1,a2)}$}
\includegraphics[width= 6cm,angle=0]{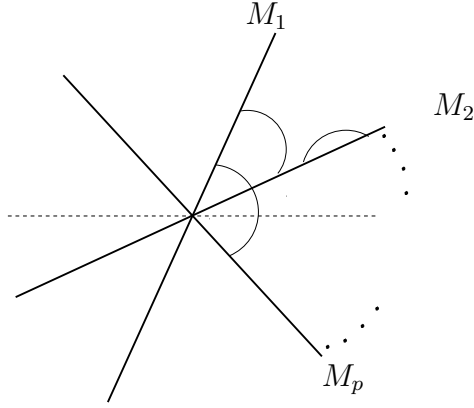}
\caption{Intersecting configuration with multiple stacks of $D3$ branes. The angles between the stacks are small but have been enlarged 
in the figure for clarity in the illustration. }
\label{intfig2}
\end{psfrags}
\end{center}
\end{figure}

Here, $c_1$ appears $M_1$ times, $c_2$ appears $M_2$ times, et cetera, such that $\sum_i M_i c_i=0$ thereby ensuring that $\tr T_D=0$. The normalization, is such that $\tr T_D^2=\f{1}{2}$. The configuration of branes corresponding to this expectation value of the scalar is
illustrated in Figure \ref{intfig2}.

There are $^pC_2 = \f{p!}{2!(p-2)!}$ off-diagonal groups of generators. The first group has $2M_1M_2$ elements, and is written next.

\begin{eqnarray}\label{multipleoffdiagonal}
T_{p+1}^{\alpha_1=1}=\f{1}{2}\left(\scalemath{0.7}{
\begin{BMAT}[7pt]{ccc:ccc:ccc:ccc}{ccc:ccc:ccc:ccc}
&&&1&0&\cdots&&& &&&\\
&\makebox[\wd0]{\large $0_{M_1\times M_1}$}&&0&0&\cdots&&\makebox[\wd0]{\large $0_{M_1\times M_3}$}&&&\makebox[\wd0]{\large $\cdots$}&\\
&&&\vdots&\vdots&\ddots&&& &&&\\
1&0&\cdots&&&&&& &&&\\
0&0&\cdots&&\makebox[\wd0]{\large $0_{M_2\times M_2}$}&&&\makebox[\wd0]{\large $0_{M_2\times M_3}$}&&&\makebox[\wd0]{\large $\cdots$}&\\
\vdots&\vdots&\ddots&&&&&& &&&\\
&&&&&&&&&&&\\
&\makebox[\wd0]{\large $0_{M_3\times M_1}$}&&&\makebox[\wd0]{\large $0_{M_3\times M_2}$}&&&\makebox[\wd0]{\large $0_{M_3\times M_3}$}&&&\makebox[\wd0]{\large $\cdots$}&\\
&&&&&&&&&&&\\
&&&&&&&&&&&\\
&\vdots&&&\vdots&&&\vdots&&&\ddots&\\
&&&&&&&&&&&\\
\end{BMAT}}\right)
~~,~~
T_{p+1}^{\alpha_1=2}=\f{1}{2}\left(\scalemath{0.7}{
\begin{BMAT}[7pt]{ccc:ccc:ccc:ccc}{ccc:ccc:ccc:ccc}
&&&-i&0&\cdots &&&&&&\\
&\makebox[\wd0]{\large $0_{M_1\times M_1}$}&&0&0&\cdots&&\makebox[\wd0]{\large $0_{M_1\times M_3}$}&&&\makebox[\wd0]{\large $\cdots$}&\\
&&&\vdots&\vdots&\ddots &&&&&&\\
i&0&\cdots&&&&&&&&&\\
0&0&\cdots&&\makebox[\wd0]{\large $0_{M_2\times M_2}$}&&&\makebox[\wd0]{\large $0_{M_2\times M_3}$}&&&\makebox[\wd0]{\large $\cdots$}&\\
\vdots&\vdots&\ddots&&&&&&&&&\\
&&&&&&&&&&&\\
&\makebox[\wd0]{\large $0_{M_3\times M_1}$}&&&\makebox[\wd0]{\large $0_{M_3\times M_2}$}&&&\makebox[\wd0]{\large $0_{M_3\times M_3}$}&&&\makebox[\wd0]{\large $\cdots$}&\\
&&&&&&&&&&&\\
&&&&&&&&&&&\\
&\vdots&&&\vdots&&&\vdots&&&\ddots&\\
&&&&&&&&&&&\\
\end{BMAT}}\right),...
\end{eqnarray}

Similarly, we can write the other groups of off-diagonal generators, $T_{p+2}^{\alpha_2}$, $T_{p+3}^{\alpha_3},...$ where $\alpha_2 = 1, 2,..., 2M_1M_3$, $\alpha_3 = 1, 2,..., 2M_2M_3$, etc. Using these, the adjoint fields can be written as
\beqa
A_{\mu} = A_{\mu}^DT_D + A^{ra_r}_{\mu}T^{a_r}_r + A_{\mu}^{(p+1)\alpha_1}T^{\alpha_1}_{p+1} + \cdots,\\
\Phi_I = \Phi_I^DT_D + \Phi_I^{ra_r}T_r^{a_r} + \Phi_I^{(p+1)\alpha_1}T^{\alpha_1}_{p+1} + \cdots .
\eeqa

%To evaluate the tree level action, we again consider the quadratic term in equation (\ref{quadterm}) and start with equation (\ref{expand2}). We note, using the forms of $T_D$ and $T^{\alpha1}_{p+1},\cdots$, that
%\beqa
%[T_{p+1}^1,T_D]=i(c_2-c_1)T_{p+1}^2~~~ \mbox{and}~~~ [T_{p+1}^2,T_D]=-i(c_2-c_1)T_{p+1}^1,
%\eeqa
%and similarly for all other pairs in each group of generators $T_{p+k}$, $k = 1,2,\cdots,^pC_2$. Using these, (\ref{expand2}) gets modified to

%\beqa\label{expand4}
%q(c_1-c_2)\left(x\pa^{\mu}\Phi_1^{(p+1)1}A_{\mu}^{(p+1)2}-x\pa^{\mu}\Phi_1^{(p+1)2}A_{\mu}^{(p+1)1}-\Phi_1^{(p+1)1}A_1^{(p+1)2}+\Phi_1^{(p+1)2}A_1^{(p+1)1}\right)+\cdots.
%\eeqa

To evaluate the tree level action, we again consider the quadratic term in equation (\ref{quadterm}) and start with equation (\ref{expand2}). We note, using the forms of $T_D$ and $T^{\alpha1}_{p+1},\cdots$, that
\beqa
[T_{p+1}^1,T_D]=i(c_2-c_1)T_{p+1}^2,~~ [T_{p+1}^2,T_D]=-i(c_2-c_1)T_{p+1}^1; \cdots.
\eeqa
Here, $\cdots$ represents all other pairs of generators in the group $T_{p+1}$. For generators $T_{p+2}$, we have
\beqa
[T_{p+2}^1,T_D]=i(c_3-c_1)T_{p+2}^2,~~ [T_{p+2}^2,T_D]=-i(c_3-c_1)T_{p+2}^1; \cdots,
\eeqa
and similarly for other groups of generators, $T_{p+3}$, $T_{p+4}$, $\cdots$, $T_{p+^pC_2}$. Using these relations, (\ref{expand2}) can be written as
\beqa\label{expand4}
&&q(c_1-c_2)\left(x\pa^{\mu}\Phi_1^{(p+1)1}A_{\mu}^{(p+1)2}-x\pa^{\mu}\Phi_1^{(p+1)2}A_{\mu}^{(p+1)1}-\Phi_1^{(p+1)1}A_1^{(p+1)2}+\Phi_1^{(p+1)2}A_1^{(p+1)1}+\cdots\right)\non
&&+q(c_3-c_1)\left(x\pa^{\mu}\Phi_1^{(p+1)1}A_{\mu}^{(p+1)2}-x\pa^{\mu}\Phi_1^{(p+1)2}A_{\mu}^{(p+1)1}-\Phi_1^{(p+1)1}A_1^{(p+1)2}+\Phi_1^{(p+1)2}A_1^{(p+1)1}+\cdots\right)\non
&&+\cdots.
\eeqa
Defining
\beqa\label{multiplet2}
\xi_k^{1}=\left(\begin{array}{c}\Phi^{(p+k)1}_1\\A^{(p+k)2}_1\\A^{(p+k)2}_2\\A^{(p+k)2}_3\end{array}\right)~;~\zeta_k^{1}=\left(\begin{array}{c}\Phi^{(p+k)1}_1\\A^{(p+k)2}_1\end{array}\right)~~~;~~~\xi_k^{'1}=\left(\begin{array}{c}\Phi^{(p+k)2}_1\\A^{(p+k)1}_1\\A^{(p+k)1}_2\\A^{(p+k)1}_3\end{array}\right)~;~\zeta_k^{'1}=\left(\begin{array}{c}\Phi^{(p+k)2}_1\\A^{(p+k)1}_1\end{array}\right)
\eeqa
(and similarly for the other pairs) for each $k = 1,2,\cdots,^pC_2$, the quadratic action for bosons is written as
\beqa\label{actionb2}
S_b&=&\int d^4 z ~\left[\sum_{s_1=1}^{M_1M_2}\left(\f{1}{2}(\xi_1^{s_1})^T{\cal O}^{1}_B\xi_1^{s_1}+\f{1}{2}(\xi_1^{'s_1})^{T}{\cal O}^{'1}_B\xi_1^{'s_1}\right)+\sum_{s_2=1}^{M_2M_3}\left(\f{1}{2}(\xi_2^{s_2})^T{\cal O}^{2}_B\xi_2^{s_2}+\f{1}{2}(\xi_2^{'s_2})^{T}{\cal O}^{'2}_B\xi_2^{'s_2}\right)+\cdots \right.\non
&~&~~~~~~~~~~~~~~~~~~~~~~~~~~~~~~~~~~~~~~~~~~~~~~~~~~~~~+\left.{\cal L}(A^{D}_{\mu},A^{1a_1}_{\mu},\cdots,A^{pa_p}_{\mu},\Phi_I,\tilde{\Phi}_J)\right],
\eeqa
where ${\cal O}^{k}_B$ is defined in (\ref{ob}) with $q$ replaced  appropriately for different pairs of stacks. Note that unlike the case of two stacks we have $^pC_2$ different factors multiplying $q$ which is reflection of the fact that the brane stacks subtend different angles with the horizontal. 

Following the computation in section \ref{treenn}, we thus have $2M_1M_2+2M_2M_3+...$ tachyons. These tachyons and the other massive modes correspond to strings that stretch between pairs of stacks as in Figure \ref{intfig2}. They transform as bi-fundamentals under pairs of $SU(M_r)$ gauge groups. The massless modes come from strings with end points on the same stack. These give rise to the gauge group $SU(M_1)\times\cdots\times SU(M_p)\times U(1)^{p-1}$.   

The eigenfunctions corresponding to the operator ${\cal O}^k_B$ are given in the appendix \ref{eigenfns}. For the present case one only needs to take care of the fact that $q$ should be replaced by $|c_i-c_j|q$ for the tachyons and the other massive modes that stretch between the $M_i$ and the $M_j$ stacks. The eigenvalues of ${\cal O}^k_B$ are then $\lambda_n^{ij}=(2n-1)|c_i-c_j|q$. Similarly for the other massive modes, the momentum modes along $x$ are given by $\gamma_n^{ij}=(2n+1)|c_i-c_j|q$. For each pair of stacks the fields can be mode expanded as in the previous sections taking care of the different factors multiplying $q$. The analysis for the tree-level fermions follow along the same lines as section \ref{treennf} with the above modifications.

We now proceed towards the analysis of one-loop two-point amplitude for the tachyons. We shall not present here the expression for the full amplitude as it is complicated and is not very illuminating. We outline how much we can carry forward the computations done in the previous sections and further discuss the new structures involved.

Due to the unbroken symmetry we only need to focus on one pair of the tachyons. Without loss of generality, let us consider the tachyon with the corresponding generator $T_{p+1}^{\alpha_1=1}$ given in (\ref{multipleoffdiagonal}). This tachyon stretches between the stacks of $M_1$ and $M_2$ branes.
Table \ref{tablecolor} summarizes the color factors associated with the one-loop two-point diagrams. The counting of these color factors is same for scalars, fermions and vectors propagating in the loop. 

The structure of the one-loop amplitude for each of the diagrams indicated in the Table \ref{tablecolor} may be summarized for the four class of diagrams (I-IV) as follows. 
\\

\noindent
I. ~~The one-loop diagram involves a four-point vertex. The massive modes propagating in the loop could either be (a) stretched between the $M_1-M_2$ stacks or (b) stretched between  $M_1-M_r$ or $M_2-M_r$ stacks. The case (a) is a configuration analogous to the the system with two stacks only. This contribution has been worked out in the previous section (for scalars). The amplitude in this case can be obtained by replacing $q$ by $q_{12}=|c_1-c_2|q$.

In case (b) the various contributions depend on the end-point of the stretched string which in this case is the stack with $M_r$ branes. For a general $r$ this is worked out in section \ref{ib} for scalars.
\\

\noindent
II. ~~The tachyon couples to the massless modes corresponding to strings that ends on $M_1$ stack or $M_2$ stack. This computation is same as the for the case of two stacks with the above replacement of $q$.
\\

\noindent
III. ~~This contribution involves a three point vertex. The modes propagating in the loop are, one massless and one massive. This is the general structure of a three point vertex that appears in the case of two stacks configuration (Figure \ref{ver3ptm1m2fig}). The contribution to the amplitude is thus same as the two stacks configuration.
\\

\noindent
IV. ~~This is a new sector that appears in the configuration consisting of more than two stacks. A typical contribution is worked out in section \ref{iv} for scalars in the loop.
\\

The full two-point one-loop amplitude can thus be schematically written as:

\beqa\label{fullmultiple}
\Sigma^2_{\mbox{\scriptsize{tachyon}}}=\left(\frac{M_1+M_2}{2}\right)\times \Sigma^2(c_1,c_2)+\sum_{r\ne 1,2}\frac{M_r}{4}\left[\mbox{I(b)}+\mbox{IV}\right](c_1,c_2,c_r)
\eeqa

where $\Sigma^2(c_1,c_2)$ has been worked out in \cite{1}.

\begin{table}[H]
\captionsetup{font=small}
\begin{center}
\begin{tabular}{m{1cm}m{4cm}cm{5cm}m{4cm}c}
\hline
& \scalebox{.9}{Loop Diagram} &&\scalebox{.9}{String modes} &\scalebox{.9}{ Color factor}&\\ 
\hline
&&&&&\\
\multirow{2}{*}{\raisebox{-3\normalbaselineskip}[0pt][0pt]{I}}&\multirow{2}{*}[-5mm]{~~~~~
\begin{psfrags}
\psfrag{c1}[][]{\scalebox{0.8}{$C$}}
\psfrag{c2}[][]{\scalebox{0.8}{$C$}}
\psfrag{c4}[][]{\scalebox{0.8}{Massive}}
\psfrag{v2}[][]{}
\includegraphics[width=2cm,angle=0]{fourpointbosonic.eps}
\end{psfrags}}
&\multirow{2}{*}{\raisebox{0\normalbaselineskip}[0pt][0pt]{(a)}}&
\begin{psfrags}
\psfrag{1}[][]{\scalebox{0.8}{$M_1$}}
\psfrag{2}[][]{\scalebox{0.8}{$M_2$}}
\psfrag{p}[][]{\scalebox{0.8}{$M_r$}}
%\psfrag{A3}[][]{$A^{(a0,a3)}/\Phi_I^{(a0,a3)}$}
%\psfrag{A2}[][]{$A^{(a1,a2)}/\Phi_I^{(a1,a2)}$}
\raisebox{0\height}{\includegraphics[width=3cm,angle=0]{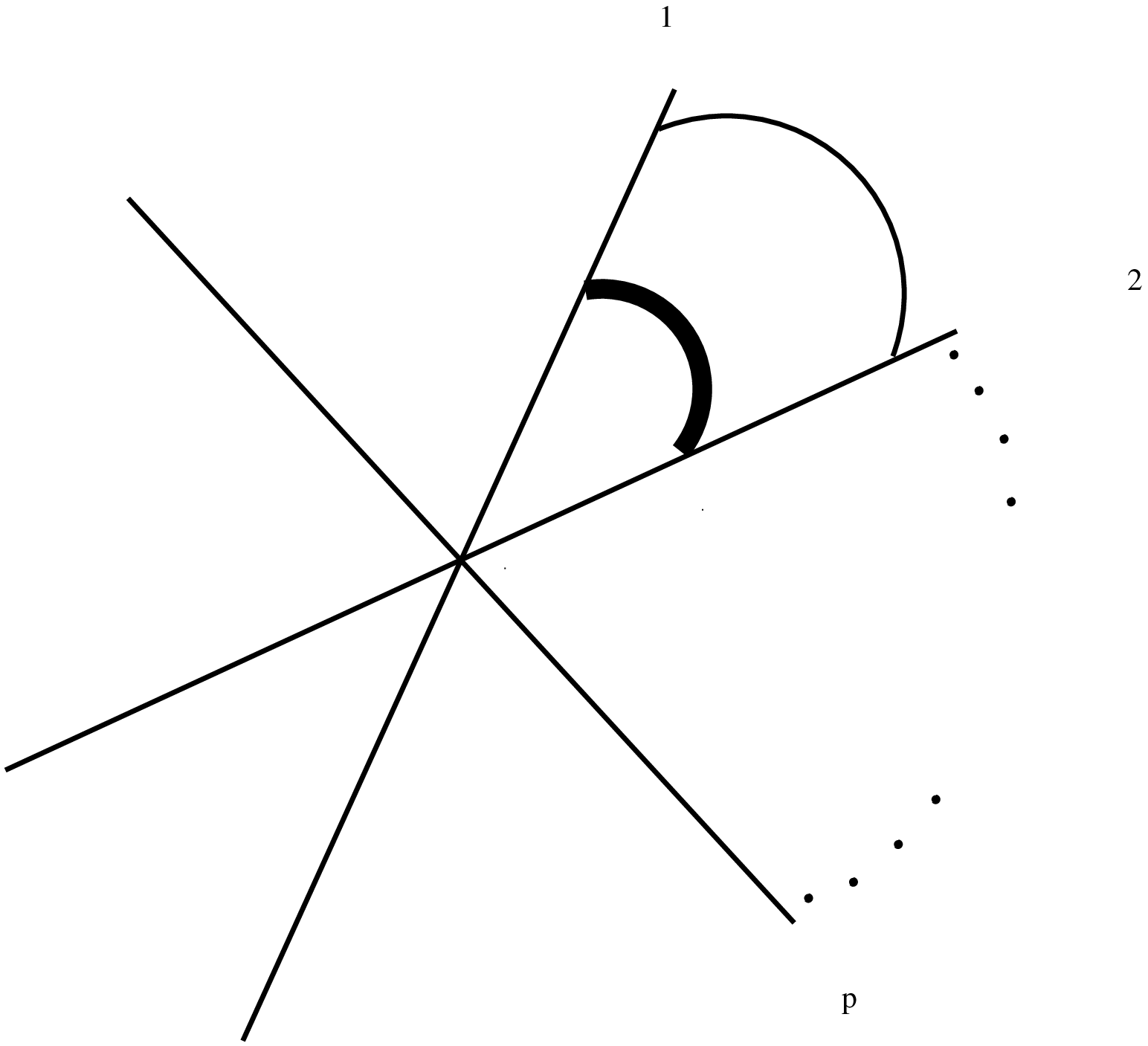}}
\end{psfrags}
&$\frac{1}{4}(M_1+M_2)$&\\[8ex]
\cline{3-5}
&&\multirow{2}{*}{\raisebox{0\normalbaselineskip}[0pt][0pt]{(b)}}&
\begin{psfrags}
\psfrag{1}[][]{\scalebox{0.8}{$M_1$}}
\psfrag{2}[][]{\scalebox{0.8}{$M_2$}}
\psfrag{p}[][]{\scalebox{0.8}{$M_r$}}
%\psfrag{A3}[][]{$A^{(a0,a3)}/\Phi_I^{(a0,a3)}$}
%\psfrag{A2}[][]{$A^{(a1,a2)}/\Phi_I^{(a1,a2)}$}
\raisebox{-1\height}{\includegraphics[width=3cm,angle=0]{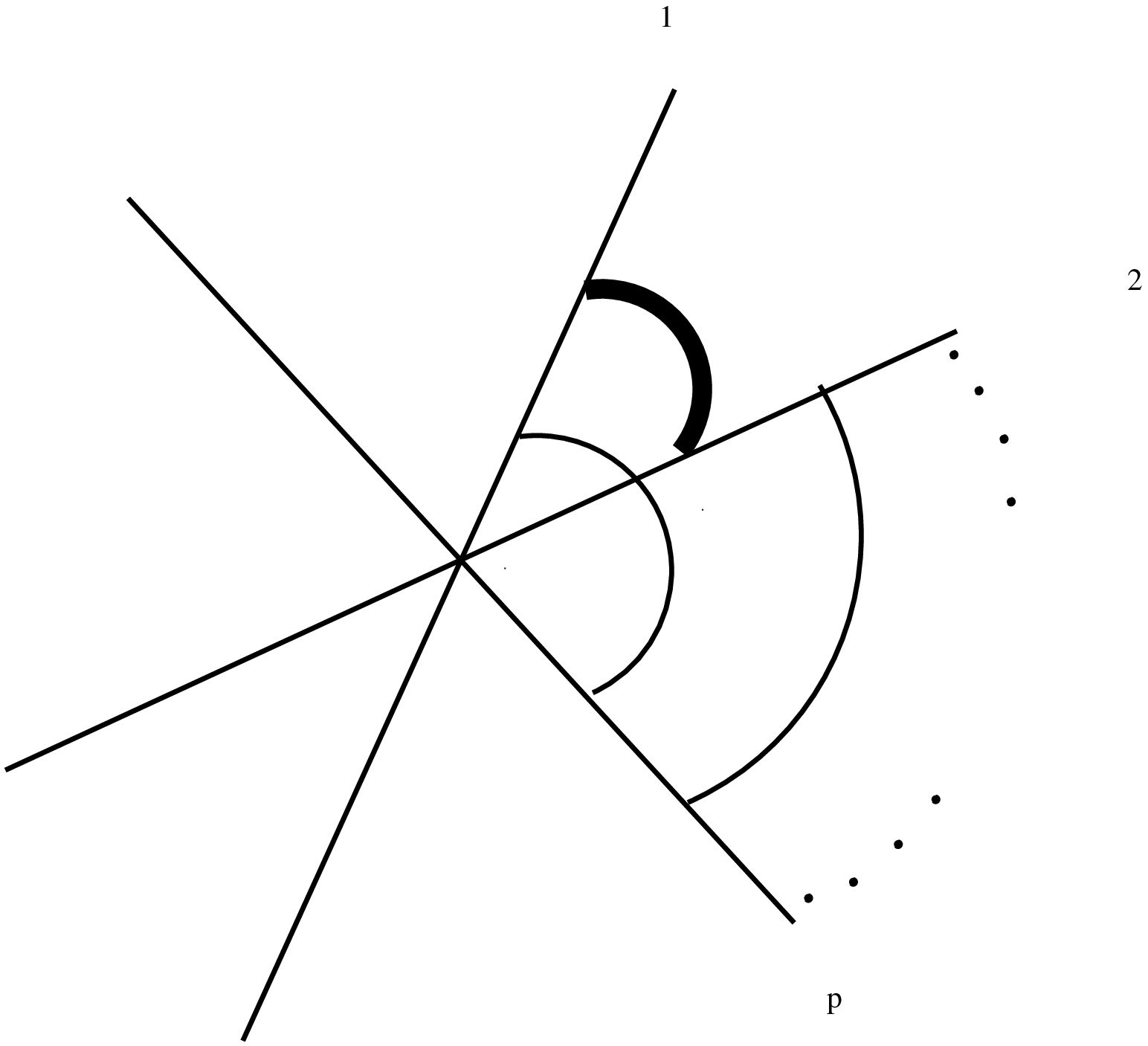}}
\end{psfrags}
&$\frac{1}{4}M_r$~;~$\frac{1}{4}M_r$&\\
\hline\\
II&~~~~~\begin{psfrags}
\psfrag{c1}[][]{\scalebox{0.8}{$C$}}
\psfrag{c2}[][]{\scalebox{0.8}{$C$}}
\psfrag{c4}[][]{\scalebox{0.8}{Massless}}
\psfrag{v2}[][]{}
\includegraphics[width=2cm,angle=0]{fourpointbosonic.eps}
\end{psfrags}
&&\begin{psfrags}
\psfrag{1}[][]{\scalebox{0.8}{$M_1$}}
\psfrag{2}[][]{\scalebox{0.8}{$M_2$}}
\psfrag{p}[][]{\scalebox{0.8}{$M_r$}}
%\psfrag{A3}[][]{$A^{(a0,a3)}/\Phi_I^{(a0,a3)}$}
%\psfrag{A2}[][]{$A^{(a1,a2)}/\Phi_I^{(a1,a2)}$}
\includegraphics[width=3cm,angle=0]{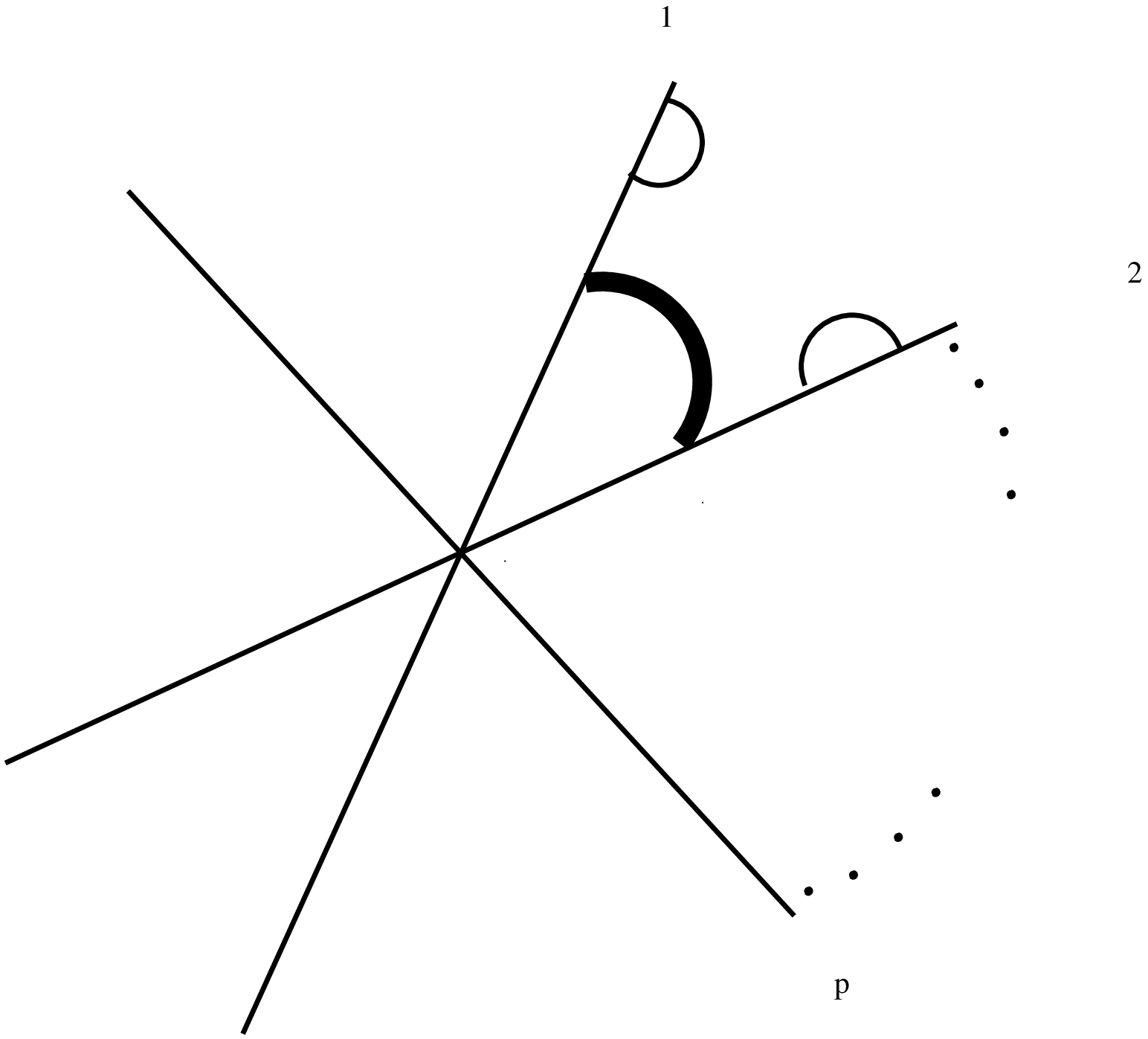}
\end{psfrags}
&$\frac{1}{4}(M_1+M_2)$&\\
\hline
III&\begin{psfrags}
\psfrag{c1}[][]{\scalebox{0.8}{$C$}}
\psfrag{c2}[][]{\scalebox{0.8}{$C$}}
\psfrag{c3}[][]{\scalebox{0.8}{Massless}}
\psfrag{c4}[][]{\scalebox{0.8}{Massive}}
\psfrag{v1}[][]{}
\psfrag{v2}[][]{}
\includegraphics[width=3.5cm,angle=0]{massless3pointf2.eps}
\end{psfrags}
&&\raisebox{-1\height}{\begin{psfrags}
\psfrag{1}[][]{\scalebox{0.8}{$M_1$}}
\psfrag{2}[][]{\scalebox{0.8}{$M_2$}}
\psfrag{p}[][]{\scalebox{0.8}{$M_r$}}
%\psfrag{A3}[][]{$A^{(a0,a3)}/\Phi_I^{(a0,a3)}$}
%\psfrag{A2}[][]{$A^{(a1,a2)}/\Phi_I^{(a1,a2)}$}
\includegraphics[width=3cm,angle=0]{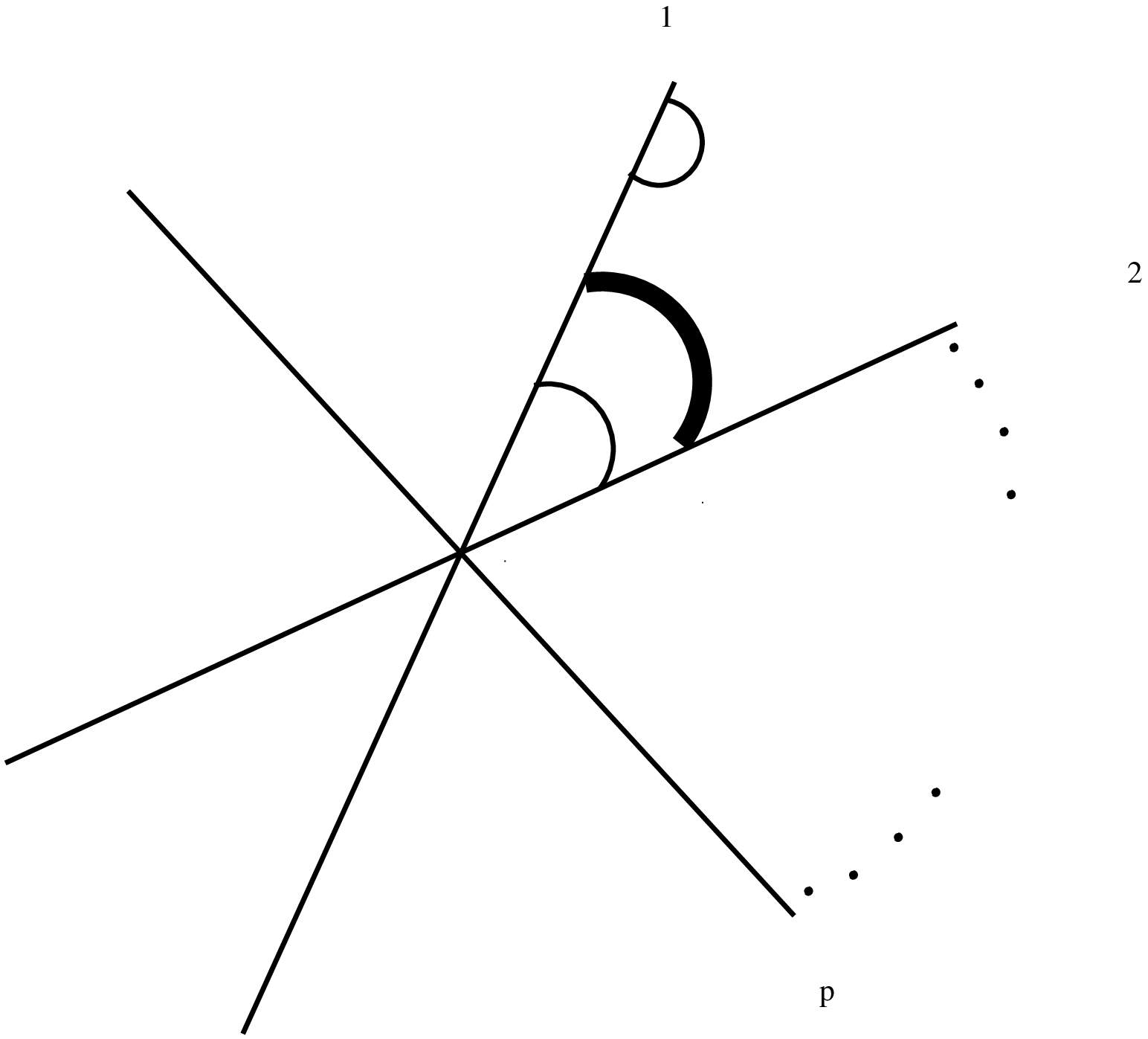}
\end{psfrags}}
&$\frac{1}{4}(M_1+M_2)$&\\[10ex]
\hline
IV&\begin{psfrags}
\psfrag{c1}[][]{\scalebox{0.8}{$C$}}
\psfrag{c2}[][]{\scalebox{0.8}{$C$}}
\psfrag{c3}[][]{\scalebox{0.8}{Massive}}
\psfrag{c4}[][]{\scalebox{0.8}{Massive}}
\psfrag{v1}[][]{}
\psfrag{v2}[][]{}
\includegraphics[width=3.5cm,angle=0]{massless3pointf2.eps}
\end{psfrags}
&&\raisebox{-1\height}{\begin{psfrags}
\psfrag{1}[][]{\scalebox{0.8}{$M_1$}}
\psfrag{2}[][]{\scalebox{0.8}{$M_2$}}
\psfrag{p}[][]{\scalebox{0.8}{$M_r$}}
%\psfrag{A3}[][]{$A^{(a0,a3)}/\Phi_I^{(a0,a3)}$}
%\psfrag{A2}[][]{$A^{(a1,a2)}/\Phi_I^{(a1,a2)}$}
\includegraphics[width=3cm,angle=0]{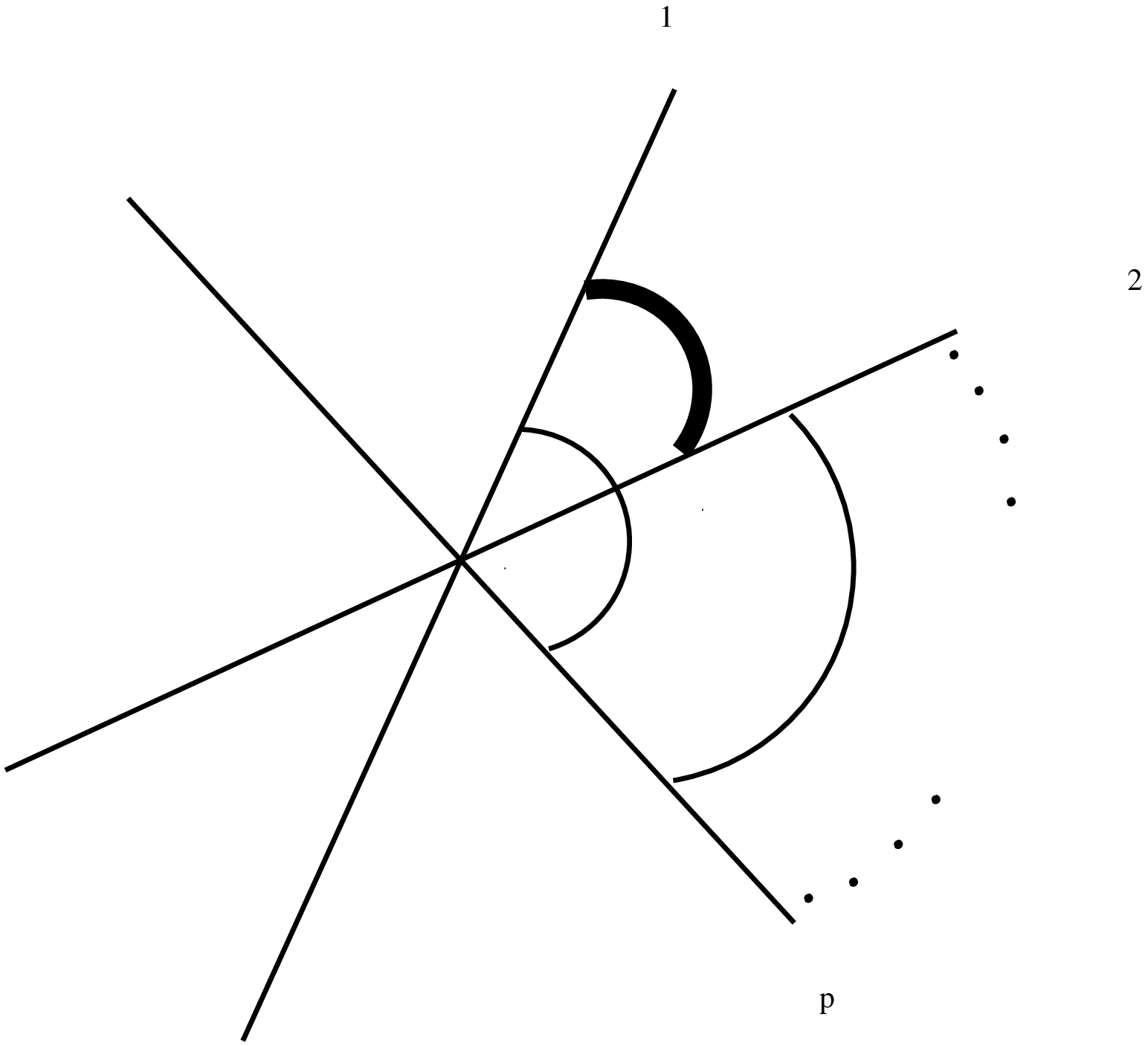}
\end{psfrags}}
&$\frac{1}{4}M_r$&\\[10ex]
\hline
\end{tabular}
\end{center}
\caption{One-loop diagrams corresponding to the two-point tachyon amplitude and their corresponding color factors are shown. In the figures representing the string modes the thick solid line corresponds to a tachyon stretched between the stacks of $M_1$ and $M_2$ branes.}
\label{tablecolor}
\end{table}

\newpage
One can check the computation of the amplitude as follows. Consider a system consisting of three stacks consisting of $M_1$, $M_2$ and $M_3$ branes. When the third stack is made to coincide with the second, one reproduces the results of the calculation for a two stacks system consisting of $M_1$ 
and $M_2+M_3$ branes. It can accordingly be seen from the Table \ref{tablecolor} that the color factors sum up to an overall value of $\frac{1}{4}(M_1+M_2+M_3)$ for the full one-loop two-point tachyon amplitude, as is the case for the configuration of two stacks of branes. 

The amplitude (\ref{fullmultiple}) is expected to be UV finite as the underlying ${\cal N}=4$ SYM theory is finite. This can be demonstrated for the one-loop amplitude by noting the following. We already have shown the cancellation of the UV contributions between the bosons and fermions in the loop for $\Sigma^2$ in \cite{1}. Further the fact that the extra new sector in (\ref{fullmultiple}) is also finite follows from the above observation of making two stacks coincident.

\subsection{Computation of I(b) for scalars}\label{ib} 

\begin{figure}[h]
\begin{center}
\begin{psfrags}
\psfrag{c1}[][]{$C$}
\psfrag{c2}[][]{$C$}
\psfrag{c4}[][]{$\Phi^a_J/\tilde{\Phi}^{a}_J$}
\psfrag{v2}[][]{$V^{I(b)}$}
\includegraphics[width= 2.5cm,angle=0]{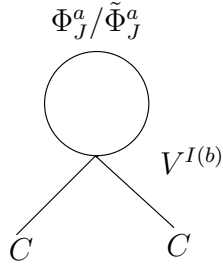}
\end{psfrags}
\caption{One loop diagram corresponding to I(b) in Table \ref{tablecolor}.}
\label{tachampboson4ptpstacks}
\end{center}
\end{figure}

In this section we compute the amplitude shown in Figure \ref{tachampboson4ptpstacks} corresponding to the configurations I(b) in Table \ref{tablecolor}.

The vertex involved corresponds to the term $ \hf \tr[\Phi_I,\Phi_J]^2 $. Contribution to the desired amplitude comes from

\beqa
(\Phi^{(p+1)1}_1)^2\Phi^{a}_J\Phi^{b}_J[T^1_{p+1},T^{a}][T^1_{p+1},T^{b}] 
\eeqa

The generators $T^{a}$ and $T^{b}$ corresponds to the massive scalar mode of the string that stretches between the stacks $M_1-M_r$ or $M_2-M_r$.

The mode expansions for the tachyon corresponding to string stretched between stacks $M_1-M_2$ is 

\beqa\label{zeta1}
\Phi^{(p+1)1}_1=N_{12}^{1/2}\int\f{d^2{\bf k}}{(2\pi\sqrt{q})^2}\sum_{m,n}\left[C(m,n,{\bf k})\phi_n(x)+
\tilde{A}^{(p+1)}_1(m,n,{\bf k})\tilde{\phi}_n(x)\right]e^{-i(\o_m\tau+{\bf k.y})}
\eeqa

\beqa
\phi_n(x) &=& {\cal N}(n)e^{- q_{12} x^2/2} \left(H_n (\sqrt{q_{12}} x) - 2 n H_{n-2} (\sqrt{q_{12}} x) \right)\non
\tilde{\phi}_n(x) &=& \tilde{{\cal N}}(n) e^{- q_{12} x^2/2} \left(H_n (\sqrt{q_{12}} x) + 2 (n-1) H_{n-2} (\sqrt{q_{12}} x) \right)
\eeqa

with $N_{ij}=\sqrt{q_{ij}}/\beta$ and $q_{ij}=q|c_i-c_j|$. The color index of the tachyon $C$ has been suppressed.

Similarly the scalar field corresponding to the string stretched between $M_1-M_r$ has the mode expansion,

\beqa\label{scmassive}
\Phi_{J}=N_{1r}^{1/2}\int\f{d^2{\bf k}}{(2\pi\sqrt{q})^2}\sum_{m,n}\Phi_{J}(m,n,{\bf k}){\cal N}^{'}(n)e^{-q_{1r}x^2/2}H_n(\sqrt{q_{1r}}x)e^{-i(\o_m\tau+{\bf k.y})}.
\eeqa

 We have also suppressed the color indices.

We now write down the vertices  which is similar to that written in the previous section.

\begin{table}[H]
\begin{center}
\begin{tabular}{lcc}
\begin{psfrags}
\psfrag{a11}[][]{\scalebox{0.8}{$\Phi_J^{a}/\tilde{\Phi}^{a}_J$}}
\psfrag{a12}[][]{\scalebox{0.8}{$\Phi_J^{b}/\tilde{\Phi}^{b}_J$}}
\psfrag{c1}[][]{\scalebox{0.8}{$C(m^{''},n^{''},{\bf k}^{''})$}}
\psfrag{c2}[][]{\scalebox{0.8}{$C(\tilde{m}^{''},\tilde{n}^{''},\tilde{{\bf k}}^{''})$}}
\psfrag{v2}[][]{$V^{I(b)}$}
\parbox[c]{3cm}{\includegraphics[width= 2.5cm,angle=0]{vertex3.eps}}
%\caption{$V_1$ , $V_2$ vertices}
\label{v1}
\end{psfrags}
&~~~&
\parbox[c]{11.8cm}{$\begin{array}{c}V^{I(b)}=-\f{N_{12}}{qg^2} F^{I(b)}(n,n^{'},n^{''},\tilde{n}^{''}) (2\pi)^2\delta^2({\bf k}+{\bf k}^{'}+{\bf k}^{''}+\tilde{{\bf k}}^{''})\delta_{m+m^{'}+m^{''}+\tilde{m}^{''}}\\
\times \tr[T^1_{p+1},T^{a}][T^1_{p+1},T^{b}]\\
F^{I(b)}(n,n^{'},n^{''},\tilde{n}^{''})=\sqrt{q_{1r}}\int dx e^{-q_{1r}x^2}\left [\phi_{n^{''}}(x)\phi_{\tilde{n}^{''}}(x)\right]\\
~~~~~~~~~~~~~~~~~~~~~~~~~~~~~~~~~~~~~~~~~~~~~~~~~~~~\left[H_n(\sqrt{q_{1r}}x)  H_{n^{'}}(\sqrt{q_{1r}}x)\right]\\
\\
\end{array}$}
\end{tabular}
\end{center}
\end{table}

Note that this vertex has a different structure than that worked out in the two stacks system due to the appearance of different $q_{ij}$ at different places.
 
The propagator for the $\Phi_J$ fields can similarly be written as

\beqa
\label{propphi4alpha1etcI}
\expect{\Phi_J^{a}(m,n,{\bf k})\Phi_J^{b}(m^{'},n^{'},{\bf k}^{'})}=qg^2\f{\delta_{m,-m^{'}}\delta_{n,n^{'}}(2\pi)^2\delta^2({\bf k}+{\bf k}^{'})\delta^{ab}}{\o_m^2+\gamma_n^{1r}+|{\bf k}|^2}.
\eeqa

Noting that the trace of the commutators of the generators evaluates to $\frac{1}{4}M_r$
the required amplitude comes out to be

\beqa
\hspace{-10mm}\left(\frac{M_r}{4}\right)N_{12} \sum_{m,n} \int \f{d^2 {\bf k}}{(2\pi\sqrt{q})^2} F^{I(b)}(n,n,0,0)\f{5}{\o_m^2 + \gamma_n^{1r}+|{\bf k}|^2}.
\eeqa

\subsection{Computation of IV for scalars}\label{iv}

\begin{figure}[h]
\begin{center}
\begin{psfrags}
\psfrag{c1}[][]{$C$}
\psfrag{c2}[][]{$C$}
\psfrag{c3}[][]{$\Phi_J^{a}/\tilde{\Phi}^{a}_J$}
\psfrag{c4}[][]{$\Phi_J^{b}/\tilde{\Phi}^{b}_J$}
\psfrag{v1}[][]{$V^{IV}$}
\psfrag{v2}[][]{$~~V^{{IV}*}$}
\includegraphics[width= 6cm,angle=0]{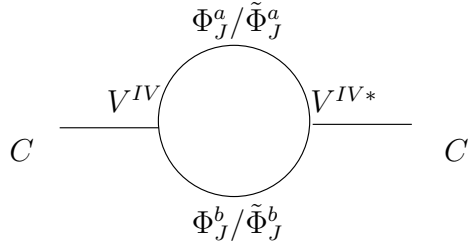}
\end{psfrags}
\caption{One loop diagram corresponding to IV in Table \ref{tablecolor}.}
\label{tachampboson3ptpstacks}
\end{center}
\end{figure}

Next, we consider the amplitude in figure \ref{tachampboson3ptpstacks} from configuration IV in Table \ref{tablecolor}.

One of the relevant terms in the action is $-2i\tr(\partial^{\mu}\Phi_I[A_{\mu},\Phi_I^b])$. Expanding the term using the generators, contribution to the requisite vertices comes from
\beqa
-2i\tr\partial_{x}\Phi_I^{a}A^{(p+1)2}_1\Phi_I^bT^a\left[T^2_{p+1},T^b\right]
\eeqa
with $T^a$ and $T^b$ having the same interpretation as in section \ref{ib}. Tachyons appearing here have mode expansion of the form
\beqa\label{zeta2}
A^{(p+1)2}_1=N_{12}^{1/2}\int\f{d^2{\bf k}}{(2\pi\sqrt{q})^2}\sum_{m,n}\left[C(m,n,{\bf k})A_n(x)+
\tilde{A}^{(p+1)2}_1(m,n,{\bf k})\tilde{A}_n(x)\right]e^{-i(\o_m\tau+{\bf k.y})}
\eeqa
where
\beqa
A_n(x) &=& {\cal N}(n)e^{- q_{12} x^2/2} \left(H_n (\sqrt{q_{12}} x) + 2 n H_{n-2} (\sqrt{q_{12}} x) \right)\non
\tilde{A}_n(x) &=& \tilde{{\cal N}}(n) e^{- q_{12} x^2/2} \left(H_n (\sqrt{q_{12}} x) - 2 (n-1) H_{n-2} (\sqrt{q_{12}} x) \right).
\eeqa
%with $N_{ij}=\sqrt{q_{ij}}/\beta$ and $q_{ij}=q|c_i-c_j|$.

In addition to above, a contribution to the amplitude in question comes from the term $\hf\tr\left[\Phi_I,\Phi_J\right]^2$ in the form
\beqa
qx\Phi^a_J\Phi^{(p+1)1}_1\Phi^b_J\tr\left(\left[T_D,T^a\right]\left[T^1_{p+1},T^b\right]\right).
\eeqa

Using the equations \ref{zeta1}, \ref{zeta2} and \ref{scmassive}, we can write the vertices as
\begin{table}[H]
\begin{center}
\begin{tabular}{lcc}
\begin{psfrags}
\psfrag{c}[][]{$C(m^{''},n^{''},{\bf k}^{''})$}
\psfrag{a1}[][]{$\Phi_J^{a}/\tilde{\Phi}_J^{a}(m,n,{\bf k})$}
\psfrag{a3}[][]{$\Phi_J^{b}/\tilde{\Phi}_J^{b}(m^{'},n^{'},{\bf k}^{'})$}
\psfrag{v4}[][]{$V^{IV}$}
\parbox[c]{5cm}{\includegraphics[width= 5 cm,angle=0]{vertex1.eps}} 
\end{psfrags}
&~~~&
\parbox[c]{10cm}{$\begin{array}{c}V^{IV}=-\f{N_{1r}^{1/2}N_{2r}^{1/2}N_{12}^{1/2}}{qg^2} \beta (2\pi)^2\delta^2({\bf k}+{\bf k}^{'}+{\bf k}^{''})\delta_{m+m^{'}+m^{''}}\\
\times \left[-2i\tr(T^a[T^2_{p+1},T^b])F_1^{IV}+2\left\{\tr\left(\left[T_D,T^a\right]\left[T^1_{p+1},T^b\right]\right)\right.\right.\\
\left.\left.~~~~~~~~~~~~~~~~~~~~~~~~~~~~~~~+(a\leftrightarrow b)\right\}F_2^{{IV}}\right](n,n',n'')\\
F_1^{IV}(n,n^{'},n^{''})=\int dx   \left[A_{n^{''}}(x)\partial_x H_n(\sqrt{q_{1r}}x)H_{n'}(\sqrt{q_{2r}}x)\right.\\
\left.-(q_{1r}x)A_{n^{''}}(x) H_n(\sqrt{q_{1r}}x)H_{n'}(\sqrt{q_{2r}}x)-(1r\leftrightarrow 2r)\right]\\
%\left. -(qx) ~ \phi_n(x) H_n(\sqrt{q_{1r}}x) H_{n'}(\sqrt{q_{2r}}x)\right]\\
e^{-q_{1r}x^2/2}e^{-q_{2r}x^2/2}{\cal N}^{'2}(n)\\
F_2^{{IV}}(n,n^{'},n^{''})=-\int dx ~ (qx) ~ \phi_{n^{''}}(x) H_n(\sqrt{q_{1r}}x) H_{n'}(\sqrt{q_{2r}}x)\\
~~~~~~~~~~~~~~~~~~~~~~~~~~~~~~~~e^{-q_{1r}x^2/2}e^{-q_{2r}x^2/2}{\cal N}^{'2}(n)%\\
%F_2^{'2} \mbox{ can be obtained by interrchanging } 1r \mbox{ and} 2r \mbox{ in } F_2^2.

\end{array}$}
\end{tabular}
\end{center}
\end{table}

Using the above written vertex and propagator from \ref{propphi4alpha1etcI}, we find that the amplitude evaluates to
%\vspace{-3cm}
\beqa
&&\hspace{0cm}\left(\f{M_r}{4}\right)\f{N_{1r}N_{2r}N_{12}}{N^2} \sum_{m,n,n^{'}} \int \f{d^2 {\bf k}}{(2\pi\sqrt{q})^2}\left\{qF_1^{IV}(0,n,n')F_1^{IV}(0,n,n') + qF_2^{{IV}}(0,n,n')F_2^{{IV}}(0,n,n')\right.\non
&&~~~~~~~~\left. + 2qF_1^{IV}(0,n,n')F_2^{{IV}}(0,n,n') \right\} \f{5}{(\o_m^2 + \gamma_n^{1r}+|{\bf k}|^2)(\o_m^2 + \gamma_{n^{'}}^{2r}+|{\bf k}|^2)}.
\eeqa

\section{Discussion and summary}\label{summary}

In this paper we have extended the analytical results for the one-loop two-point amplitude for the tachyon to the case of intersecting stacks of branes. The setup of the present computation follows along the lines of \cite{2} and \cite{1}. The intersecting configuration, consisting of two stacks of branes, corresponds to breaking of $SU(M_1+M_2)$ gauge symmetry to $SU(M_1)\times SU(M_2)\times U(1)$. In the Yang-Mills approximation this is achieved by turning on an expectation value of a scalar. A more general configuration consisting of multiple stacks has also been discussed. The tachyons in these configurations correspond to open strings stretching between two stacks of branes and transform as bi-fundamental representations under two gauge groups corresponding to the two stacks. 

We have shown that for two intersecting stacks of $D3$ branes the two-point amplitude is equal to the one obtained in \cite{1} times additional color factors from the unbroken gauge symmetry. In this paper we have analyzed the tachyon mass as a function of temperature. Due to the complicated nature of the full amplitude this has to be done numerically. However, it is clear from the previous studies of the amplitude as a function of temperature in \cite{1} and \cite{2}, that a critical temperature $T_c$ exists when tachyon becomes massless. The unbroken gauge symmetry implies that the critical temperature is same for all the tachyons in the two-stacks system. For more than two stacks, there are multiple scales. 

The tachyon mass-squared is given by $m^2_{\mbox{\tiny{tachyon}}}=\frac{\partial^2}{\partial |C|^2}V(|C|^2)|_{|C|=0}$, where $C$ and $C^*$ are a pair of tachyons charged under the unbroken $U(1)$ (we have suppressed the other indices on $C$). $V(|C|^2)$ denotes the effective potential. The present calculation computes the coefficient of the quadratic term, that is $|C|^2$, in $V(|C|^2)$. This coefficient is given by $m^{2({\mbox{\tiny{tree-level}}})}_{\mbox{\tiny{tachyon}}}+\Sigma_{\mbox{\tiny{tachyon}}}$. 

Now consider a two-stacks system. The tree level tachyon mass is given by $m^{2({\mbox{\tiny{tree-level}}})}_{\mbox{\tiny{tachyon}}}=-q_{12}/(2qg^2)=-|c_1-c_2|/(2g^2)$, with $|c_1-c_2|=\sqrt{M_1+M_2}/\sqrt{2M_1M_2}$. For $\beta^2 q<<1$, the one loop correction has the form

\beqa\label{estimate}
\Sigma_{\mbox{\tiny{tachyon}}}\sim \left(\frac{M_1+M_2}{2}\right)\left[|c_1-c_2|x_0+\f{x}{q\beta^2}\right],
\eeqa

where $x_0$ is the zero temperature contribution and $x$ is the temperature independent part of the one-loop amplitude for single intersecting branes computed in \cite{1}. We can now estimate the critical temperature for the two-stacks system. Using equation \ref{estimate}, this is given by $T_c/\sqrt{q_{12}}\sim 1/(g\sqrt{(M_1+M_2)/2})$. For $M_1=M_2=1$, the configuration reduces to a pair of intersecting branes where $T_c/\sqrt{q}\sim 1/g$ \cite{1}.

At temperatures above this critical temperature, the intersecting configuration is stable. Such a transition is also expected in the holographic BCS model \cite{3}. The BCS instability in this holographic model is mapped to the instability of intersecting $D8$ branes in $D4$ brane background. The present computation involves a simplified configuration of intersecting stacks in flat space. Nevertheless it captures the existence of a critical temperature, which is an essential feature of phase transition. As a next step, one would now wish to address questions related to the order of transition and the properties of the condensed phase. This requires the knowledge of the effective potential. This can be explored using nonperturbative techniques as discussed in \cite{Sen} (see \cite{Daniel} and \cite{Majumdar} for related studies). However, it needs to be seen whether the perturbative techniques discussed here may also be reliably used.

\vspace{1cm}
\noindent
{\large {\bf Acknowledgments}}\\
We would like to thank Sudipto Paul Chowdhury for collaboration at initial stages of this work and Balachandran Sathiapalan for reading this manuscript. V.S. acknowledges CSIR, India, for support through SRF grant 09/045(1355)/2014-EMR-I.

\appendix

\section{${\cal N}=4$ SYM in $4D$}\label{dimred}

The ${\cal N}=4$ SYM in $4D$ action is given by (see \cite{dimred1}, \cite{dimred2} and appendix B of \cite{1} for a review of dimensional reduction from $10D$ ${\cal N}=1$ SYM),

\beqa\label{n4action}
S_4=S^1_{4}+S^2_4
\eeqa

with

\beqa\label{actionboson}
S_{4}^1=\f{1}{g^2}\mbox{tr}\int d^{4}x \left[-\f{1}{2}F_{\mu\nu}F^{\mu\nu}-D_{\mu}\Phi_ID^{\mu}\Phi_I-D_{\mu}\tilde{\Phi}_ID^{\mu}\tilde{\Phi}_I+\f{1}{2}\left([\Phi_I,\Phi_J]^2+[\tilde{\Phi}_I,\tilde{\Phi}_J]^2+2[\Phi_I,\tilde{\Phi}_J]^2\right)\right]\non
\eeqa

\beqa\label{actionfermion}
S_{4}^2=\f{1}{g^2}\mbox{tr}\int d^{4}x \left[-i\bar{\lambda}_k\gamma^{\mu}D_{\mu}\lambda_k+\bar{\lambda}_k[(\alpha_{kl}^I\Phi_I+\beta_{kl}^I\gamma^5\tilde{\Phi}),\lambda_l]\right]
\eeqa

where the $\alpha$ and $\beta$ matrises satisfy
\beqa
\{\alpha^I,\alpha^J\}=\{\beta^I,\beta^J\}=-2\delta^{IJ} ~~~[\alpha^I,\beta^J]=0
\eeqa

and are written as

\beqa
\alpha^1=\left(\begin{array}{cc}0&\sigma^1\\-\sigma^1&0\end{array}\right)~~~\alpha^2=\left(\begin{array}{cc}0&\sigma^3\\-\sigma^3&0\end{array}\right)~~~\alpha^3=\left(\begin{array}{cc}i\sigma^2&0\\0&i\sigma^2\end{array}\right)
\eeqa

\beqa
\beta^1=\left(\begin{array}{cc}0&i\sigma^2\\i\sigma^2&0\end{array}\right)~~~\beta^2=\left(\begin{array}{cc}0&1\\-1&0\end{array}\right)~~~\beta^3=\left(\begin{array}{cc}-i\sigma^2&0\\0&i\sigma^2\end{array}\right)
\eeqa

The 4D $\gamma$-matrices are

\beqa
\gamma^0=\left(\begin{array}{cc}0&i\sigma^2\\i\sigma^2&0\end{array}\right)~~~\gamma^1=\left(\begin{array}{cc}0&-\sigma_3\\-\sigma_3&0\end{array}\right)~~~\gamma^2=\left(\begin{array}{cc}0&-\sigma^1\\-\sigma^1&0\end{array}\right)~~~\gamma^3=\left(\begin{array}{cc}\mathbb{I}_2&0\\0&-\mathbb{I}_2 \end{array}\right)
\eeqa

\section{Eigenfunctions}\label{eigenfns}

\subsection{Bosons}
In this section we list the eigenfunctions of the operator ${\cal O}^{11}_B$. These have also been discussed elaborately in \cite{2,1}.

\noindent

{\bf Eigenfunctions:},

\beqa\label{cfunctions}
\zeta_n(x)=\left(\begin{array}{c}
\phi_n(x)\\
-A_n(x)
\end{array}\right) ~~~~\tilde{\zeta}_n(x)=\left(\begin{array}{c}
\tilde{\phi}_n(x)\\
-\tilde{A}_n(x)
\end{array}\right)
\eeqa

where 

\begin{gather}
\label{bosegnfunc}
A_n(x) =   {\cal N}(n)e^{- q x^2/2} \left(H_n (\sqrt{q} x) + 2 n H_{n-2} (\sqrt{q} x) \right) \\
\phi_n(x) = {\cal N}(n)e^{- q x^2/2} \left(H_n (\sqrt{q} x) - 2 n H_{n-2} (\sqrt{q} x) \right)
\end{gather} 

\begin{gather}
\label{zerobosegnfunc}
\tilde{A}_n(x) =  \tilde{{\cal N}}(n) e^{- q x^2/2} \left(H_n (\sqrt{q} x) - 2 (n-1) H_{n-2} (\sqrt{q} x) \right) \\
\tilde{\phi}_n(x) = \tilde{{\cal N}}(n) e^{- q x^2/2} \left(H_n (\sqrt{q} x) + 2 (n-1) H_{n-2} (\sqrt{q} x) \right)
\end{gather}

\noindent
{\bf Normalizations:} 

\beqa
{\cal N}(n)&=&\f{1}{\sqrt{\sqrt{\pi} 2^n (4 n^2-2n) (n-2)!}}\non
\tilde{{\cal N}}(n)&=&\frac{1}{\sqrt{\sqrt{\pi} 2^n (4 n-2)(n-1)!}}
\eeqa

\noindent
{\bf Eigenvalues:}

The eigenvalues corresponding to $\zeta_n(x)$ are $(2n-1)q$ and those corresponding to $\tilde{\zeta}_n(x)$ are all zero. Thus the spectrum in the latter case is completely degenerate. In the non-zero eigenvalue sector we do not have normalizable eigenfunctions corresponding to $n=1$. However unlike this sector, in the zero eigenvalue sector we have normalizable eigenfunction for $n=1$, which is simply $H_1(\sqrt{q} x)$ but there is 
no normalizable eigenfunctions for $n=0$ in this sector. 

\noindent
{\bf Orthogonality conditions:}

\beqa\label{ortho1}
\sqrt{q}\int dx \zeta^{\dagger}_n(x)\zeta_{n^{'}}(x)=\delta_{n,n^{'}} ~~~~\sqrt{q}\int dx \tilde{\zeta}^{\dagger}_n(x)\tilde{\zeta}_{n^{'}}(x)=\delta_{n,n^{'}}
\eeqa

\beqa\label{ortho2}
\sqrt{q}\int dx \zeta^{\dagger}_n(x)\tilde{\zeta}_{n^{'}}(x)=0 \mbox{~~ for all~~$n$~and~$n^{'}$~~}.
\eeqa

Similarly the eigenfunctions of the operator ${\cal O}^{'11}_B$ are simply $\zeta^{'}=(\phi_n(x), A_n(x))$, 
and $\tilde{\zeta}^{'}=(\tilde{\phi}_n(x), \tilde{A}_n(x))$ with eigenvalues $(2n-1)q$ and $0$ respectively.

\subsection{Fermions}

In this section we list the eigenfunctions of ${\cal O}_F^x$ and $\tilde{{\cal O}}_F^x$ 

\noindent
{\bf Eigenfunctions:}

\beqa
\left(\begin{array}{c}L_n(x)\\R_n(x)\end{array}\right) ~~\mbox{and}~~ \left(\begin{array}{c}L_n(x)\\-R_n(x)\end{array}\right)
\eeqa

\begin{equation}
\label{fermsolnLR}
\begin{split}
&L_n(x) =  {\cal N}_F e^{- \frac{q x^2}{2}}\left(- \frac{i}{\sqrt{2n}} H_{n}(\sqrt{q} x) 
+  H_{n-1} (\sqrt{q} x)\right)\\
&R_n(x) = {\cal N}_F e^{- \frac{q x^2}{2}}\left(- \frac{i}{\sqrt{2n}} H_{n}(\sqrt{q} x) 
-  H_{n-1} (\sqrt{q} x)\right). 
\end{split}
\end{equation}

$H_n(\sqrt{q}x)$ are the Hermite Polynomials. 

{\bf Normalization and Eigenvalues:}

\beqa
{\cal N}_F=\f{1}{\sqrt{\sqrt{\pi}2^{n+1} (n-1)!}} 
\eeqa

The eigenvalues are $-i\sqrt{\lambda^{'}_n}=-i\sqrt{2nq}$.

{\bf Relations:}

\begin{gather}
\label{fermorthocon}
\sqrt{q}\int dx~\psi^{\dagger}_n(x) \psi_{n^{'}}(x) = \sqrt{q}\int dx \left(L^*_n(x) 
L_{n^{'}}(x) + R^*_n(x) R_{n^{'}}(x)\right) = \delta_{n,n^{'}}.\\
\label{fermreln1}
\sqrt{q}\int dx~L^*_n(x) L_{n^{'}}(x) = \sqrt{q}\int dx~R^*_n(x) R_{n^{'}}(x) = \hf \delta_{n,n^{'}}\\
\label{fermreln2}
\sqrt{q}\int dx~\psi^{T}_n(x) \psi_{n^{'}}(x)=\sqrt{q}\int dx \left(L_n(x) L_{n^{'}}(x) 
+ R_n(x) R_{n^{'}}(x)\right) = 0\\
\label{fermreln3}
\sqrt{q}\int dx~\psi^{\dagger}_n(x) \psi^{*}_{n^{'}}(x)=\sqrt{q}\int dx \left(L^{*}_n(x) L^{*}_{n^{'}}(x) 
+ R^{*}_n(x) R^{*}_{n^{'}}(x)\right) = 0    
\end{gather}

\end{document}